\documentclass[extra, mreferee]{gji}

\usepackage{epsfig, color}
\usepackage{amssymb, amsmath}
\usepackage{setspace}

\title[Green's function representations]
 {Green's function representations for Marchenko imaging without up/down decomposition}

\author[Wapenaar et al.]
{\small Kees Wapenaar$^1$, Roel Snieder$^2$, Sjoerd de Ridder$^3$  and Evert Slob$^1$\\
$^1$Department of Geoscience and Engineering, Delft University of Technology, P.O. Box 5048, 2600 GA Delft, \\The Netherlands\\
$^2$Center for Wave Phenomena, Colorado School of Mines, Golden CO 80401, USA\\
$^3$School of Earth and Environment, University of Leeds, Leeds, LS2 9JT, United Kingdom}
\pagerange{1--99}
\volume{999}
\pubyear{2021}

\begin{document}
\begin{spacing}{2.0}
\label{firstpage}

\maketitle

\begin{summary}
{\small Marchenko methods are based on integral representations which express Green's functions for virtual sources and/or receivers in the subsurface
in terms of the reflection response at the surface. An underlying assumption is that inside the medium the wave field can be decomposed into downgoing and upgoing 
waves and that evanescent waves can be neglected. We present a new derivation of Green's function representations  which circumvents
these assumptions, both for the acoustic and the elastodynamic situation. These representations form the basis for research into new Marchenko methods which have
the potential to handle refracted and evanescent waves and to more accurately image steeply dipping reflectors.
\mbox{}\\}
\end{summary}

\begin{keywords}
{\small Controlled source seismology, Seismic interferometry, Wave scattering and diffraction}
\end{keywords}

\section{Introduction}\label{sec1}

Marchenko redatuming, imaging, monitoring and multiple elimination are all derived from
 integral representations which express Green's functions for virtual sources or receivers in the subsurface
in terms of the reflection response at the surface
\citep{Ravasi2016GJI, Staring2018GEO, Jia2018GEO, Lomas2019GEO, Mildner2019GEO, Brackenhoff2019SE, Zhang2020GJI, Elison2020GJI, Reinicke2020GEO}.
These representations, in turn, are derived from reciprocity theorems for one-way wave fields \citep{Slob2014GEO, Wapenaar2014GEO}, building on ideas presented by \citet{Broggini2012EJP}.
Marchenko methods deal with internal multiples in a data-driven way and have the potential to solve large-scale 3D imaging and multiple elimination problems
\citep{Pereira2019SEG, Staring2020GP, Ravasi2021GEO}. Of course Marchenko methods have also limitations.
One of the limitations is caused by the fact that the one-way reciprocity theorems require that the wave field in the subsurface region of interest 
can be decomposed into downgoing and upgoing fields. Moreover, one
of these reciprocity theorems (the correlation-type theorem) is based on the assumption that evanescent waves can be neglected. 
These assumptions complicate the imaging of steep flanks and exclude a proper treatment of refracted waves and evanescent waves tunnelling through high velocity layers. 

To address some of the limitations, \cite{Kiraz2021JASA} propose a Marchenko method without decomposition inside the medium, assuming the input data are acquired on a closed boundary.
On the other hand, for reflection data on a single horizontal boundary, a first step has been made towards a Marchenko method that deals with evanescent waves  \citep{Wapenaar2020GJI}. This method
is restricted to horizontally layered media and uses wave field decomposition inside the medium.

In this paper we derive more general Green's function representations 
which do not rely on wave field decomposition in the subsurface and which hold for an arbitrarily inhomogeneous medium below a single horizontal acquisition boundary.
These representations form a starting point for new research on Marchenko methods which circumvent several of the present limitations.
 \cite{Diekmann2021PRR} independently investigate the same problem, 
but without specifying a focusing condition for their focusing function $f$ and requiring a time-symmetric source function for $f$. 
 Our derivation follows a different approach, using an explicit focusing condition and requiring no source function for our focusing function $f$.
Moreover, we derive several forms of Green's function representations, including one for the homogeneous Green's function between a virtual source and a virtual receiver in the subsurface. 
We also derive elastodynamic versions of these representations. 

This paper is restricted to the derivation of the Green's function representations; a discussion of their application in new Marchenko methods is beyond the scope of this paper. 

\section{Acoustic wave field representation}\label{sec2}

We consider a lossless acoustic medium, consisting of a homogeneous isotropic upper half-space and an arbitrary inhomogeneous anisotropic lower half-space, separated
by a horizontal surface ${{\partial\mathbb{D}}_R}$. Coordinates in the medium are denoted by ${\bf x}=({\bf x}_{\rm H},x_3)$, with ${\bf x}_{\rm H}=(x_1,x_2)$ denoting the horizontal coordinates and $x_3$ the depth coordinate
(the positive $x_3$-axis is pointing downward). The horizontal surface ${{\partial\mathbb{D}}_R}$ is defined at $x_3=x_{3,R}$ (in the next section we choose this as the surface at which
seismic acquisition takes place).
The medium parameters of the lower half-space $x_3>x_{3,R}$ are the compressibility $\kappa({\bf x})$ and the mass density tensor $\rho_{jk}({\bf x})$. At the micro scale (much smaller than the wavelength of the 
acoustic field) the mass density is isotropic. However, small-scale heterogeneities of the isotropic mass density, for example caused by fine-layering, may manifest themselves 
as effective anisotropy at the scale of the wavelength \citep{Schoenberg83JASA}. The mass density tensor is symmetric, that is, $\rho_{jk}({\bf x})=\rho_{kj}({\bf x})$.
The parameters of the upper half-space $x_3<x_{3,R}$ are the constant compressibility $\kappa=\kappa_0$ and the constant isotropic mass density $\rho_{jk}=\delta_{jk}\rho_0$, 
where $\delta_{jk}$ is the Kronecker delta function. 
The propagation velocity of the upper half-space is $c_0=(\kappa_0\rho_0)^{-1/2}$.  At ${{\partial\mathbb{D}}_R}$ 
we choose the same constant isotropic medium parameters as in the upper half-space.

\begin{figure}
\vspace{-.3cm}
\centerline{\epsfysize=10 cm \epsfbox{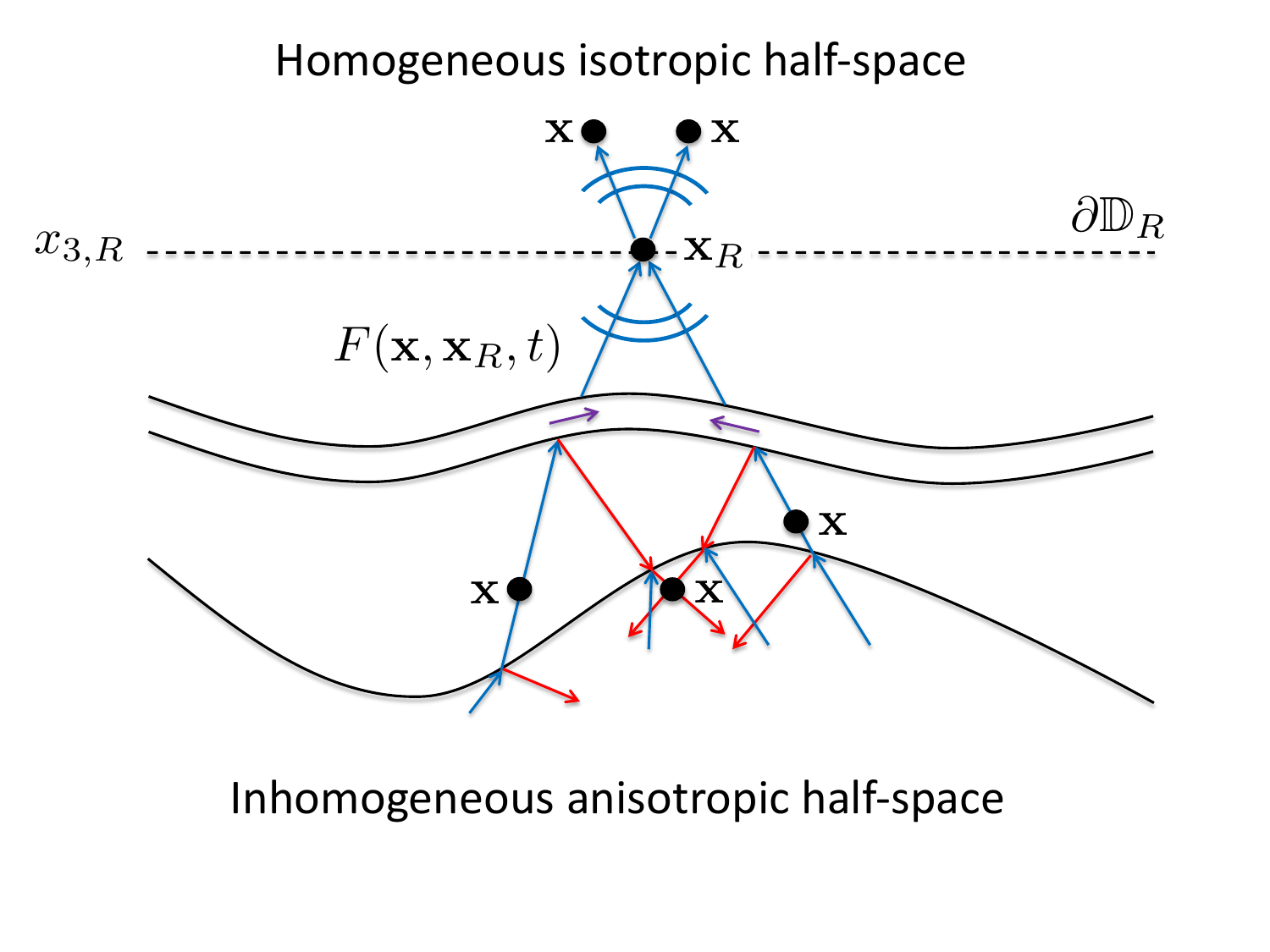}}
\vspace{-1.cm}
\caption{Illustration of the focusing function $F({\bf x},{\bf x}_R,t)$, which focuses at ${\bf x}_R$. 
For ${\bf x}$ at and above ${{\partial\mathbb{D}}_R}$ it is purely upgoing. For ${\bf x}$ in the  half-space below ${{\partial\mathbb{D}}_R}$ 
it is a complex wave field. The near-horizontal arrows in the thin layer illustrate tunnelling evanescent waves.
}\label{Figure1}
\end{figure}

The basic equations for acoustic wave propagation are the linearized equation of motion
\begin{eqnarray}
\rho_{jk}\partial_tv_k+\partial_jp=0\label{eqbeq1}
\end{eqnarray}
and the linearized deformation equation
\begin{eqnarray}
\kappa\partial_tp+\partial_iv_i=q,\label{eqbeq2}
\end{eqnarray}
respectively. Here $p({\bf x},t)$ is the space (${\bf x}$) and time ($t$) dependent acoustic pressure, $v_i({\bf x},t)$ the  particle velocity and $q({\bf x},t)$ a source in terms
of volume-injection rate density. Operator $\partial_i$ stands for differentiation in the $x_i$-direction. Lower-case subscripts (except $t$) take on the values 1, 2 and 3,
and the summation convention applies to repeated subscripts. Operator $\partial_t$ stands for differentiation with respect to time.
We introduce the specific volume tensor $\vartheta_{ij}({\bf x})$ as the inverse of the mass density tensor, with $\vartheta_{ij}\rho_{jk}=\delta_{ik}$. 
Applying the operator $\partial_i\vartheta_{ij}$ to equation (\ref{eqbeq1}),
operator $\partial_t$ to equation (\ref{eqbeq2}), and subtracting the two equations yields the acoustic wave equation
\begin{eqnarray}
\partial_i(\vartheta_{ij}\partial_jp)-\kappa\partial_t^2p=-\partial_t q.\label{eqwe}
\end{eqnarray}
We introduce a focusing function $F({\bf x},{\bf x}_R,t)$, in which ${\bf x}_R=({\bf x}_{{\rm H},R},x_{3,R})$ denotes the position of a focal point at ${{\partial\mathbb{D}}_R}$, see Figure \ref{Figure1}.
For fixed ${\bf x}_R$ and variable ${\bf x}$ and $t$, this focusing function is a solution of wave equation (\ref{eqwe}) for the source-free situation, hence, for $q=0$.
We define the focusing condition  as
\begin{eqnarray}
F({\bf x},{\bf x}_R,t)|_{x_3=x_{3,R}}&=&\delta({\bf x}_{\rm H}-{\bf x}_{{\rm H},R})\delta(t),\label{eq2a}
\end{eqnarray}
and further demand that $F({\bf x},{\bf x}_R,t)$ is purely upgoing at ${{\partial\mathbb{D}}_R}$ and in the homogeneous isotropic upper half-space.
Note that $F({\bf x},{\bf x}_R,t)$ is similar, but not identical, to the focusing function $f_2({\bf x},{\bf x}_R,t)$ introduced in  \citet{Wapenaar2014GEO}.
We come back to this in section \ref{sec3.2}.

We define the temporal Fourier transform of a space- and time-dependent function $u({\bf x},t)$ as
\begin{eqnarray}\label{eqA11}
u({\bf x},\omega)=\int_{-\infty}^\infty u({\bf x},t)\exp(i\omega t){\rm d}t,
\end{eqnarray}
where $\omega$ is the angular frequency and $i$ the imaginary unit. The integral is taken from $t=-\infty$ to $t=\infty$ to account for non-causal functions, such as
the focusing function $F({\bf x},{\bf x}_R,t)$ for which no causality condition is implied (hence, in general it can be non-zero for positive and negative time).
With this transform, wave equation (\ref{eqwe}) transforms to
\begin{eqnarray}
{\cal L}p=i\omega q,\label{eqweom}
\end{eqnarray}
with
\begin{eqnarray}
{\cal L}=\partial_i\vartheta_{ij}\partial_j+\omega^2\kappa.
\end{eqnarray}
The  focusing function $F({\bf x},{\bf x}_R,\omega)$ obeys in the frequency domain the  wave equation 
\begin{eqnarray}
{\cal L}F=0, \label{eqF0}
\end{eqnarray}
the  focusing condition
\begin{eqnarray}
F({\bf x},{\bf x}_R,\omega)|_{x_3=x_{3,R}}&=&\delta({\bf x}_{\rm H}-{\bf x}_{{\rm H},R}),\label{eq4}
\end{eqnarray}
and it is upgoing at and above ${{\partial\mathbb{D}}_R}$.
We discuss a representation for a wave field $p({\bf x},\omega)$, which may have sources in the upper half-space above ${{\partial\mathbb{D}}_R}$, but which obeys the source-free wave equation ${\cal L}p=0$ 
for $x_3\ge x_{3,R}$.
In the lower half-space we express $p({\bf x},\omega)$ as a superposition of  mutually independent wave fields that obey the same source-free wave equation as $p({\bf x},\omega)$ 
for $x_3\ge x_{3,R}$.
For this purpose, we choose the focusing functions $F({\bf x},{\bf x}_R,\omega)$ and $F^*({\bf x},{\bf x}_R,\omega)$ 
(the asterisk denotes complex conjugation, which corresponds to time-reversal in the time domain).
To be more specific, we express $p({\bf x},\omega)$ as
\begin{eqnarray}
p({\bf x},\omega)&=&\int_{{{\partial\mathbb{D}}_R}} F({\bf x},{\bf x}_R,\omega)a({\bf x}_R,\omega){\rm d}{\bf x}_R 
+\int_{{{\partial\mathbb{D}}_R}} F^*({\bf x},{\bf x}_R,\omega)b({\bf x}_R,\omega){\rm d}{\bf x}_R,
\nonumber\\&&\hspace{8cm}
\mbox{for}\quad x_3\ge x_{3,R}. \label{eq12again}
\end{eqnarray}
Here $a({\bf x}_R,\omega)$ and $b({\bf x}_R,\omega)$ are as yet undetermined coefficients, which depend on the position ${\bf x}_R$ at ${{\partial\mathbb{D}}_R}$. 
In Appendix \ref{AppA1} we formulate  boundary conditions for the acoustic pressure and the vertical component of the particle velocity 
at ${{\partial\mathbb{D}}_R}$, from which we solve $a({\bf x}_R,\omega)$ and $b({\bf x}_R,\omega)$. We thus obtain
\begin{eqnarray}
p({\bf x},\omega)&=&\int_{{{\partial\mathbb{D}}_R}} F({\bf x},{\bf x}_R,\omega)p^-({\bf x}_R,\omega){\rm d}{\bf x}_R 
+\int_{{{\partial\mathbb{D}}_R}} F^*({\bf x},{\bf x}_R,\omega)p^+({\bf x}_R,\omega){\rm d}{\bf x}_R,\nonumber\\
&&\hspace{8cm}\mbox{for}\quad x_3\ge x_{3,R},  \label{eq12}
\end{eqnarray}
where $p^-({\bf x}_R,\omega)$ and $p^+({\bf x}_R,\omega)$ represent the upgoing ($-$) and downgoing ($+$) parts, respectively, of $p({\bf x}_R,\omega)$ for ${\bf x}_R$ at ${{\partial\mathbb{D}}_R}$.
These upgoing and downgoing fields are pressure-normalized, meaning that $p^-+p^+=p$ at and above ${{\partial\mathbb{D}}_R}$. 
Below  ${{\partial\mathbb{D}}_R}$ we only consider the total (undecomposed) wave field $p$.

\begin{figure}
\centerline{\epsfysize=6 cm \epsfbox{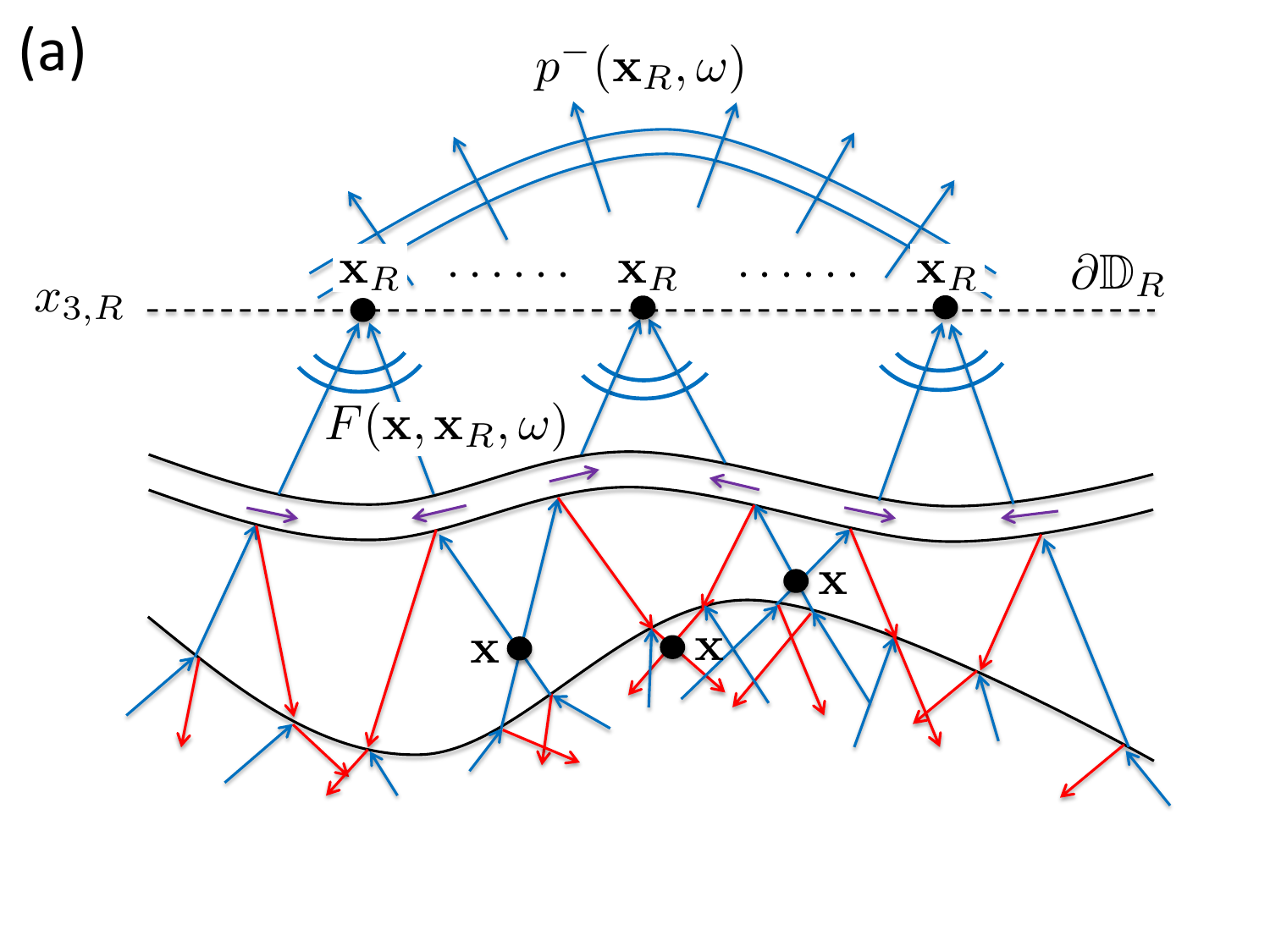}\epsfysize=6 cm\epsfbox{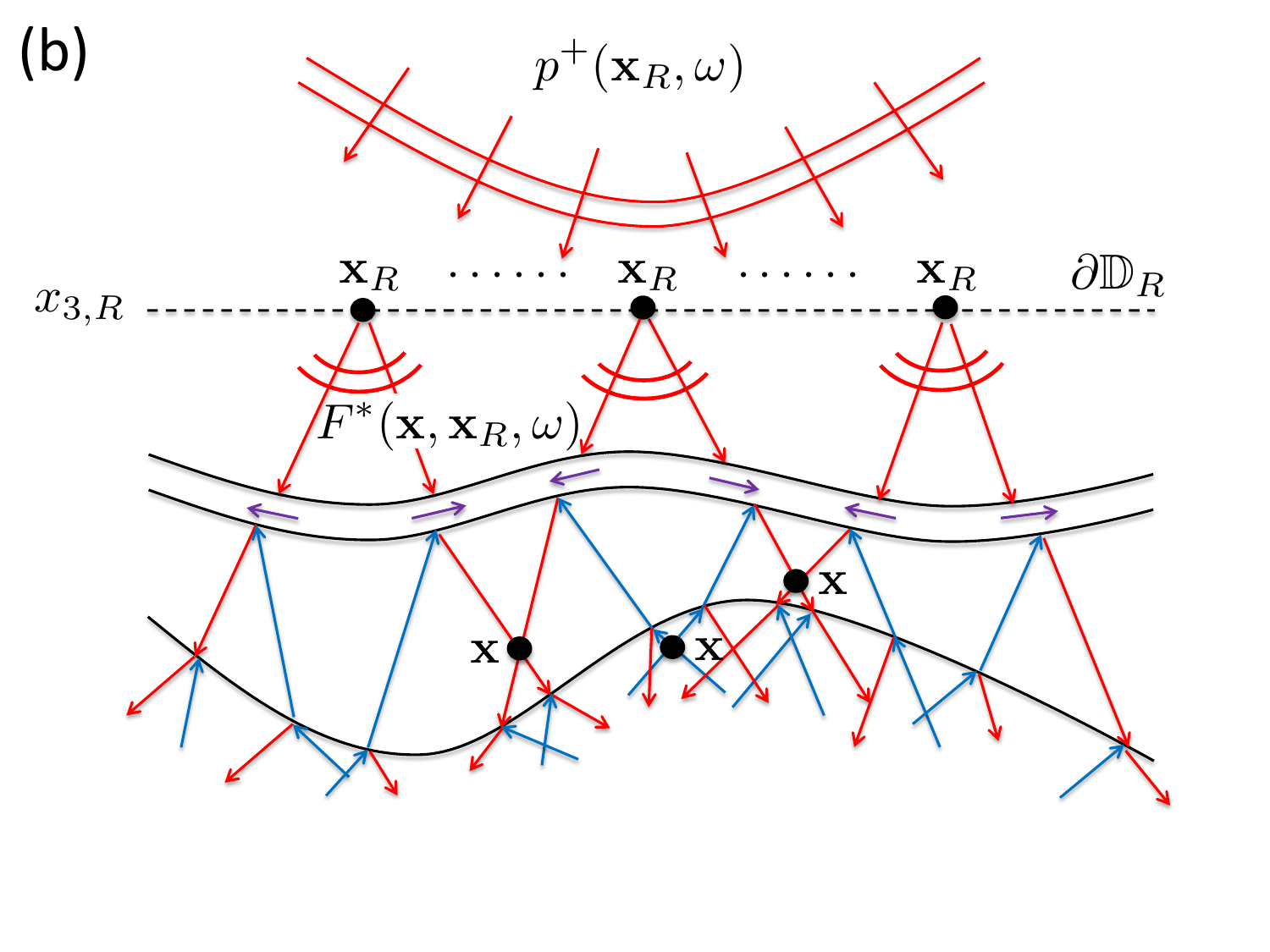}}
\vspace{-1.cm}
\caption{Explanation of the first (a) and second (b) integral in equation (\ref{eq12}) in terms of Huygens' principle. 
The focusing functions $F({\bf x},{\bf x}_R,\omega)$ and $F^*({\bf x},{\bf x}_R,\omega)$ propagate the  fields $p^-({\bf x}_R,\omega)$ and 
$p^+({\bf x}_R,\omega)$ from all ${\bf x}_R$ at the surface ${{{\partial\mathbb{D}}_R}}$ to  ${\bf x}$ in the lower half-space.
The superposition of these propagated fields gives the field $p({\bf x},\omega)$ for any ${\bf x}$ in the lower half-space.
}\label{FigA2}
\end{figure}

For an intuitive explanation of the right-hand side of 
equation (\ref{eq12}) we refer to Figure \ref{FigA2}. First we consider the second integral, which is illustrated in Figure \ref{FigA2}b. The downgoing field $p^+({\bf x}_R,\omega)$
is incident from the homogeneous upper half-space to ${{\partial\mathbb{D}}_R}$. The complex conjugate focusing function $F^*({\bf x},{\bf x}_R,\omega)$ propagates this downgoing field from
${\bf x}_R$ at ${{\partial\mathbb{D}}_R}$ to ${\bf x}$ in the lower half-space and the integral superposes the contributions from all ${\bf x}_R$ at ${{\partial\mathbb{D}}_R}$. 
Next we consider the first integral of equation (\ref{eq12}), which is illustrated in Figure \ref{FigA2}a.  Here the upgoing field $p^-({\bf x}_R,\omega)$ is backpropagated by the focusing function $F({\bf x},{\bf x}_R,\omega)$ 
from all ${\bf x}_R$ at ${{\partial\mathbb{D}}_R}$ to ${\bf x}$ in the lower half-space. The sum of the two integrals gives the wave field $p({\bf x},\omega)$ in the lower half-space.
This is a modified form of Huygens' principle, with the focal points ${\bf x}_R$ denoting the positions of the secondary sources at ${{\partial\mathbb{D}}_R}$, and with the
forward and backward propagating focusing functions replacing the Green's functions in the usual form of Huygens' principle. 
Note that the two integrals  cannot be separately associated with $p^+({\bf x},\omega)$ and $p^-({\bf x},\omega)$ in the lower half-space;
only the sum of the two integrals gives the total field $p({\bf x},\omega)$, according to equation (\ref{eq12}).

The underlying assumption in the derivation of  equation (\ref{eq12}) is that evanescent waves can be neglected at ${{\partial\mathbb{D}}_R}$. 
Hence, it only holds for waves that are propagating at ${{\partial\mathbb{D}}_R}$, having
a horizontal slowness ${\bf s}$ obeying
\begin{eqnarray}
|{\bf s}|\le 1/c_0, \quad\mbox{at}\,\,{{\partial\mathbb{D}}_R}.\label{eqevan}
\end{eqnarray}
This implies that the foci (and hence the secondary sources) at ${{\partial\mathbb{D}}_R}$ are not ideal delta functions (as formulated by equation (\ref{eq4})) 
but band-limited versions of delta functions.
Note that ignoring evanescent waves at ${{\partial\mathbb{D}}_R}$ does not imply that evanescent waves are not accounted for inside the inhomogeneous medium below ${{\partial\mathbb{D}}_R}$.
For example, in an isotropic horizontally layered medium with depth-dependent velocity $c(x_3)$, 
in which the horizontal slowness is independent of depth, 
waves that are propagating at ${{\partial\mathbb{D}}_R}$  become evanescent when they reach a depth at which $1/c(x_3)<|{\bf s}|\le 1/c_0$.
In section \ref{sec3.4} we  show with a numerical example that equation (\ref{eq12}) indeed accounts for such evanescent waves. 
Although for laterally varying media we cannot formulate a similar precise condition for waves becoming evanescent, it is still true that equation (\ref{eq12}) holds 
for  evanescent waves inside the medium, as long as they are related to propagating waves at the surface, as formulated by equation (\ref{eqevan}).

Note that we previously derived a representation similar to equation (\ref{eq12}) with heuristic arguments, 
and used it as the starting point for deriving the Marchenko method \citep{Wapenaar2013PRL}.
However, further on in that derivation we applied up/down decomposition to the wave field at an artificial internal boundary in the lower half-space 
and we neglected evanescent waves throughout space.
In the following derivations we  avoid up/down decomposition in the lower half-space and evanescent waves are only neglected at ${{\partial\mathbb{D}}_R}$.
From equation (\ref{eq12}) we derive Green's function representations for the full wave field at any point ${\bf x}$ in the subsurface, expressed in terms of the reflection response at the 
surface.

\section{Acoustic Green's function representations}

\subsection{Representation for the  acoustic dipole Green's function}\label{sec3.1}

We introduce the Green's function $G({\bf x},{\bf x}_S,t)$ as a solution of equation (\ref{eqwe}) for an impulsive monopole source of volume-injection rate density at ${\bf x}_S$, hence
\begin{eqnarray}
\partial_i(\vartheta_{ij}\partial_jG)-\kappa\partial_t^2G=-\delta({\bf x}-{\bf x}_S)\partial_t\delta(t).\label{eqweG}
\end{eqnarray}
We demand that $G$ is the causal solution of this equation, hence $G({\bf x},{\bf x}_S,t)=0$ for $t<0$. Note that $G$ obeys source-receiver reciprocity, i.e., $G({\bf x},{\bf x}_S,t)=G({\bf x}_S,{\bf x},t)$.
In the frequency domain, $G({\bf x},{\bf x}_S,\omega)$ obeys the following wave equation
\begin{eqnarray}
{\cal L} G=i\omega\delta({\bf x}-{\bf x}_S).\label{eqweGom}
\end{eqnarray}
We  choose ${\bf x}_S=({\bf x}_{{\rm H},S},x_{3,S})$ in the upper half-space, at a vanishing distance $\epsilon$ above ${{\partial\mathbb{D}}_R}$, hence, $x_{3,S}=x_{3,R}-\epsilon$.
We define a dipole-source response as
\begin{eqnarray}
\Gamma({\bf x},{\bf x}_S,\omega)=-\frac{2}{i\omega\rho_0}\partial_{3,S} G({\bf x},{\bf x}_S,\omega),\label{eq57a}
\end{eqnarray}
where  $\partial_{3,S}$ denotes differentiation with respect to the source coordinate $x_{3,S}$. 
For ${\bf x}$ at ${{\partial\mathbb{D}}_R}$ (i.e., just below the source level) we have for the downgoing part
\begin{eqnarray}
\Gamma^+({\bf x},{\bf x}_S,\omega)|_{x_3=x_{3,R}}&=&\delta({\bf x}_{\rm H}-{\bf x}_{{\rm H},S}).\label{eq4G}
\end{eqnarray}
We define the reflection response $R({\bf x}_R,{\bf x}_S,\omega)$ of the medium below ${{\partial\mathbb{D}}_R}$ as the upgoing part of the dipole-source response $\Gamma({\bf x}_R,{\bf x}_S,\omega)$, with ${\bf x}_R$ at ${{\partial\mathbb{D}}_R}$, hence
\begin{eqnarray}
R({\bf x}_R,{\bf x}_S,\omega)&=&\Gamma^-({\bf x}_R,{\bf x}_S,\omega)\nonumber\\
&=&-\frac{2}{i\omega\rho_0}\partial_{3,S} G^-({\bf x}_R,{\bf x}_S,\omega)\nonumber\\
&=&-\frac{2}{i\omega\rho_0}\partial_{3,S} G^{\rm s}({\bf x}_R,{\bf x}_S,\omega),\label{eq4Ra}
\end{eqnarray}
where superscript ${\rm s}$ stands for scattered.
Substituting  
$p({\bf x},\omega)=\Gamma({\bf x},{\bf x}_S,\omega)$ and $p^\pm({\bf x}_R,\omega)=\Gamma^\pm({\bf x}_R,{\bf x}_S,\omega)$ 
into equation (\ref{eq12}), using equations (\ref{eq4G}) and (\ref{eq4Ra}), gives
\begin{eqnarray}
\Gamma({\bf x},{\bf x}_S,\omega)&=&\int_{{{\partial\mathbb{D}}_R}} F({\bf x},{\bf x}_R,\omega)R({\bf x}_R,{\bf x}_S,\omega){\rm d}{\bf x}_R
+F^*({\bf x},{\bf x}_S,\omega), \quad\mbox{for}\quad x_3 \ge x_{3,R}. \label{eq56}
\end{eqnarray}
This is a representation for the dipole response $\Gamma({\bf x},{\bf x}_S,\omega)$ at virtual receiver position ${\bf x}$ anywhere in the half-space below ${{\partial\mathbb{D}}_R}$,
expressed in terms of the reflection response $R({\bf x}_R,{\bf x}_S,\omega)$ at ${{\partial\mathbb{D}}_R}$. It is similar to our earlier derived Green's function representations for the Marchenko method,
but here it has been derived without applying decomposition in the lower half-space. 
It only excludes the contribution from waves that are evanescent at ${{\partial\mathbb{D}}_R}$.
Another difference with our earlier representations is that the Green's function on the left-hand side is a dipole response instead of a monopole response. We address this in section \ref{sec3.2}.

It may be counterintuitive that in equation (\ref{eq56}) we use focusing functions $F({\bf x},{\bf x}_R,\omega)$ and $F^*({\bf x},{\bf x}_S,\omega)$, 
with their focal points ${\bf x}_R$ and ${\bf x}_S$ situated at the surface ${{\partial\mathbb{D}}_R}$.
This is different from the focusing functions in the classical representations for the Marchenko method, which have their focal points
in the subsurface.
For an intuitive explanation of the right-hand side of equation (\ref{eq56}) we refer again to Figure \ref{FigA2}, this time with $p^+({\bf x}_R,\omega)$ and $p^-({\bf x}_R,\omega)$ 
replaced by $\delta({\bf x}_{{\rm H},R}-{\bf x}_{{\rm H},S})$ and $R({\bf x}_R,{\bf x}_S,\omega)$, respectively. In a similar way as in section \ref{sec2}, the
focusing functions propagate the downgoing source field at ${\bf x}_S$ and the upgoing reflection response at ${\bf x}_R$ from 
${{\partial\mathbb{D}}_R}$ to ${\bf x}$ in the lower half-space, with the focal points acting again as secondary sources in a modified form of Huygens' principle. The sum
of the two integrals gives $\Gamma({\bf x},{\bf x}_S,\omega)$.

\subsection{Representation for the  acoustic monopole Green's function}\label{sec3.2}

In this section we turn equation (\ref{eq56})  into a representation for the monopole Green's function $G({\bf x},{\bf x}_S,\omega)$.
To this end we introduce a modified focusing function $f({\bf x},{\bf x}_R,\omega)$ via
\begin{eqnarray}
F({\bf x},{\bf x}_R,\omega)&=&\frac{2}{i\omega\rho_0}\partial_{3,R} f({\bf x},{\bf x}_R,\omega),\label{eq71}
\end{eqnarray}
where  $\partial_{3,R}$ denotes differentiation with respect to $x_{3,R}$.
According to equations (\ref{eqF0}), (\ref{eq4}) and (\ref{eq71}), $f({\bf x},{\bf x}_R,\omega)$ obeys the wave equation
\begin{eqnarray}
{\cal L}f=0,\label{eqwef}
\end{eqnarray}
the focusing condition
\begin{eqnarray}
\partial_{3,R} f({\bf x},{\bf x}_R,\omega)|_{x_3=x_{3,R}}&=&\frac{i\omega\rho_0}{2}\delta({\bf x}_{\rm H}-{\bf x}_{{\rm H},R}),\label{eqfoc2}
\end{eqnarray}
and it is upgoing at and above ${{\partial\mathbb{D}}_R}$.
Equation (\ref{eq71}) implies
\begin{eqnarray}
F^*({\bf x},{\bf x}_S,\omega)&=&-\frac{2}{i\omega\rho_0}\partial_{3,S} f^*({\bf x},{\bf x}_S,\omega).\label{eq72}
\end{eqnarray}
Substituting equations  (\ref{eq57a}), (\ref{eq4Ra}), (\ref{eq71}) and (\ref{eq72}) into equation (\ref{eq56}),
applying source-receiver reciprocity to the scattered Green's function 
and dropping the operation $-\frac{2}{i\omega\rho_0}\partial_{3,S}$ from all terms gives
\begin{eqnarray}
&&\hspace{-0.7cm}G({\bf x},{\bf x}_S,\omega)=
\frac{2}{i\omega\rho_0}\int_{{{\partial\mathbb{D}}_R}} \{\partial_{3,R} f({\bf x},{\bf x}_R,\omega)\}G^{\rm s}({\bf x}_S,{\bf x}_R,\omega){\rm d}{\bf x}_R+f^*({\bf x},{\bf x}_S,\omega),\nonumber\\
&&\hspace{8cm}\mbox{for}\quad x_3 \ge x_{3,R}. \label{eq56h}
\end{eqnarray}
We transfer the 
operator $\partial_{3,R}$  from $f$ to $G^{\rm s}$, which is accompanied with a sign change (see Appendix \ref{AppA2}).  Using
the definition of $R$ in equation (\ref{eq4Ra}) (with $\partial_{3,S}$ replaced by $\partial_{3,R}$ and with ${\bf x}_R$ and ${\bf x}_S$ interchanged on both sides of the equation) this yields
\begin{eqnarray}
G({\bf x},{\bf x}_S,\omega)&=&\int_{{{\partial\mathbb{D}}_R}} f({\bf x},{\bf x}_R,\omega)R({\bf x}_S,{\bf x}_R,\omega){\rm d}{\bf x}_R
+f^*({\bf x},{\bf x}_S,\omega), \quad\mbox{for}\quad x_3 \ge x_{3,R}. \label{eq56b}
\end{eqnarray}
This is the main result of this paper. We discuss a number of aspects of this representation.
\begin{itemize}
\item Equation  (\ref{eq56b}) has the same form as equation (13) in \citet{Wapenaar2014GEO}, with $f_2$ in that paper replaced by $f$.
 Using $\partial_{3,R} f({\bf x},{\bf x}_R,\omega)=-\partial_3 f({\bf x},{\bf x}_R,\omega)$ for $x_3=x_{3,R}$ (i.e., at the boundary of the homogeneous upper half-space), equation (\ref{eqfoc2}) 
 can be written as
\begin{eqnarray}
\partial_3 f({\bf x},{\bf x}_R,\omega)|_{x_3=x_{3,R}}&=&-\frac{i\omega\rho_0}{2}\delta({\bf x}_{\rm H}-{\bf x}_{{\rm H},R}).\label{eqfoc2b}
\end{eqnarray}
This is the same focusing condition as was defined for $f_2({\bf x},{\bf x}_R,\omega)$.  An important difference between $f$ and $f_2$ is the medium
in which these focusing functions are defined. Focusing function $f_2$ is defined in a truncated version of the actual medium,
where the medium below some depth level is replaced by a homogeneous medium. It is assumed that up/down decomposition is possible at the 
truncation level. On the other hand, focusing function  $f$ in equation (\ref{eq56b}) is defined in the actual medium (similar as $F$ in Figure \ref{Figure1}).

Moreover, the derivation in \cite{Wapenaar2014GEO} of the representation 
is different: in that paper we start with decomposed focusing functions $f_1^+({\bf x},{\bf x}_A,\omega)$ and $f_1^-({\bf x},{\bf x}_A,\omega)$ in
the  truncated medium, with ${\bf x}_A$ being a focal point at the truncation depth.
Next, we derive representations for decomposed Green's functions $G^+({\bf x}_A,{\bf x}_S,\omega)$ and $G^-({\bf x}_A,{\bf x}_S,\omega)$ and combine the two into a single representation 
for $G({\bf x}_A,{\bf x}_S,\omega)=G^+({\bf x}_A,{\bf x}_S,\omega)+G^-({\bf x}_A,{\bf x}_S,\omega)$, using the relation
$f_2({\bf x}_A,{\bf x}_R,\omega)=f_1^+({\bf x}_R,{\bf x}_A,\omega)-\{f_1^-({\bf x}_R,{\bf x}_A,\omega)\}^*$ (note the different order of coordinates in $f_1$ and $f_2$).
The latter relation is only valid when evanescent waves 
can be neglected at the truncation level inside the medium. In our current approach we  do not make use of decomposition at a truncation level
inside the medium and we avoid the approximate relation $f_2=f_1^+-\{f_1^-\}^*$.
The  only requirement for $f({\bf x},{\bf x}_R,\omega)$ is that it obeys equations (\ref{eqwef}) and (\ref{eqfoc2}).
Hence, representation (\ref{eq56b}) gives the  full wave field at  the virtual receiver position ${\bf x}$ inside the medium, 
including multiply reflected, refracted and evanescent waves. It only excludes the contribution from waves that are evanescent at ${{\partial\mathbb{D}}_R}$, see 
the condition formulated by equation (\ref{eqevan}).

\item Using another approach, also without applying decomposition inside the medium, \cite{Diekmann2021PRR}  derive an equation of the same form as equation (\ref{eq56b}), but without specifying
a focusing condition for $f$. Their focusing function obeys a wave equation with a non-zero time-symmetric source function which is not explicitly specified.
The derivation of equation (\ref{eq56b}) in the current paper uses an explicit focusing condition (equation \ref{eqfoc2}) and does not require a source function for $f$.

\item
Equation  (\ref{eq56b}) forms a starting point for deriving the Marchenko method. By applying an inverse Fourier transform we obtain 
\begin{eqnarray}
&&G({\bf x},{\bf x}_S,t)-f({\bf x},{\bf x}_S,-t)=
\int_{{{\partial\mathbb{D}}_R}}{\rm d}{\bf x}_R\int_{-\infty}^t f({\bf x},{\bf x}_R,t')R({\bf x}_S,{\bf x}_R,t-t'){\rm d} t',\label{eq56btime}
\end{eqnarray}
for  $x_3 \ge x_{3,R}$. The Marchenko method is based on the separability in time of $G({\bf x},{\bf x}_S,t)$ and $f({\bf x},{\bf x}_S,-t)$.
 For  horizontal plane waves in 1D media \citep{Burridge80WM, Broggini2012EJP} and for point-source responses at limited horizontal distances $|{\bf x}_{\rm H}-{\bf x}_{{\rm H},S}|$ in 
 moderately inhomogeneous 3D media \citep{Wapenaar2013PRL},
these functions only overlap at $t=t_{\rm d}$, which is the time of the direct arrival of the Green's function. This minimum overlap in time 
allows the construction of a time-windowed version of equation (\ref{eq56btime}) with  $G({\bf x},{\bf x}_S,t)$ suppressed and with $f({\bf x},{\bf x}_S,-t)$ almost completely preserved 
(this is the 3D Marchenko equation). From this equation the focusing function $f({\bf x},{\bf x}_S,t)$ can be resolved, given its direct arrival and the reflection response $R({\bf x}_S,{\bf x}_R,t)$.
In essence the separability of the Green's function
and the time-reversed focusing function has been the underlying assumption of all implementations of the  Marchenko method.
This assumption excludes, among others, the treatment of refracted waves, which may arrive prior to the direct arrival of the Green's function and interfere with the time-reversed focusing function.

Since we have argued that the representations of equations  (\ref{eq56b}) and (\ref{eq56btime}) hold for refracted and evanescent waves,
it is opportune to start new research on Marchenko methods which exploit the generality of these representations.
Care should be taken to account for the overlap in time of  the Green's function
and the time-reversed focusing function, particularly when dealing with refracted waves.
 A further discussion of the development of new Marchenko methods is beyond the scope of this paper.

\item 
Equation (\ref{eq56b}) is, in principle, suited to retrieve the Green's function $G({\bf x},{\bf x}_S,\omega)$ for ${\bf x}$ anywhere in the lower half-space. However, a single type of Green's function
 is not a sufficient starting point  for imaging.
In the classical approach to Marchenko imaging, the downgoing and upgoing parts of the Green's function are retrieved, from which a reflection image can be obtained, either by 
a deconvolution \citep{Wapenaar2014GEO, Broggini2014GEO} or a correlation method \citep{Behura2014GEO}. In the full-wavefield approach, we need at  least one other type of field at ${\bf x}$,
next to $G({\bf x},{\bf x}_S,\omega)$, which represents the acoustic pressure field at ${\bf x}$ in response to a volume-injection rate source at ${\bf x}_S$. 
To this end we introduce a Green's function $G_i^v({\bf x},{\bf x}_S,\omega)$ which, for $i$=1, 2, 3,  stands for the three components of the particle velocity field at ${\bf x}$.
From the Fourier transform of equation (\ref{eqbeq1})  we derive that the particle velocity $v_i$ can be expressed in terms of the acoustic pressure as
$v_i=\frac{1}{i\omega}\vartheta_{ij}\partial_jp$. 
Similarly, we relate $G_i^v$ to $G$ via
\begin{eqnarray}
G_i^v({\bf x},{\bf x}_S,\omega)=\frac{1}{i\omega}\vartheta_{ij}({\bf x})\partial_jG({\bf x},{\bf x}_S,\omega).\label{eqvp}
\end{eqnarray}
Hence, when $G({\bf x},{\bf x}_S,\omega)$ is available on a sufficiently dense grid, $G_i^v({\bf x},{\bf x}_S,\omega)$ can be obtained via equation (\ref{eqvp}).
Alternatively, $G_i^v({\bf x},{\bf x}_S,\omega)$ can be obtained from a modified version of the representation for $G({\bf x},{\bf x}_S,\omega)$. 
Applying the operation $\frac{1}{i\omega}\vartheta_{ij}\partial_j$ to both sides of equation (\ref{eq56b}) yields
\begin{eqnarray}
G_i^v({\bf x},{\bf x}_S,\omega)&=&\int_{{{\partial\mathbb{D}}_R}} h_i({\bf x},{\bf x}_R,\omega)R({\bf x}_S,{\bf x}_R,\omega){\rm d}{\bf x}_R
-h_i^*({\bf x},{\bf x}_S,\omega), \quad\mbox{for}\quad x_3 \ge x_{3,R},\label{eq56c}
\end{eqnarray}
with
\begin{eqnarray}
h_i({\bf x},{\bf x}_R,\omega)=\frac{1}{i\omega}\vartheta_{ij}({\bf x})\partial_jf({\bf x},{\bf x}_R,\omega).
\end{eqnarray}
The  Green's functions $G({\bf x},{\bf x}_S,\omega)$ and $G_i^v({\bf x},{\bf x}_S,\omega)$ together  provide sufficient information for imaging.
For example, one could decompose the field into incident and scattered waves in any desired direction, say in a direction perpendicular to a local interface
 \citep{Yoon2006EG, Liu2011GEO, Holicki2019GP}, and use these fields as input for imaging.

\end{itemize}

\subsection{Representation for the homogeneous acoustic Green's function}\label{sec3.3}

The representations in sections \ref{sec3.1} and \ref{sec3.2} give the response  to a source at ${\bf x}_S$, 
observed by a virtual receiver at ${\bf x}$ inside the medium.
Here we modify the representation of equation (\ref{eq56b}), to create the response at the surface to 
a virtual source inside the medium. After that, we show how to obtain the response to this virtual source at a virtual receiver inside the medium.

\begin{figure}
\vspace{-.5cm}
\centerline{\epsfysize=7 cm \epsfbox{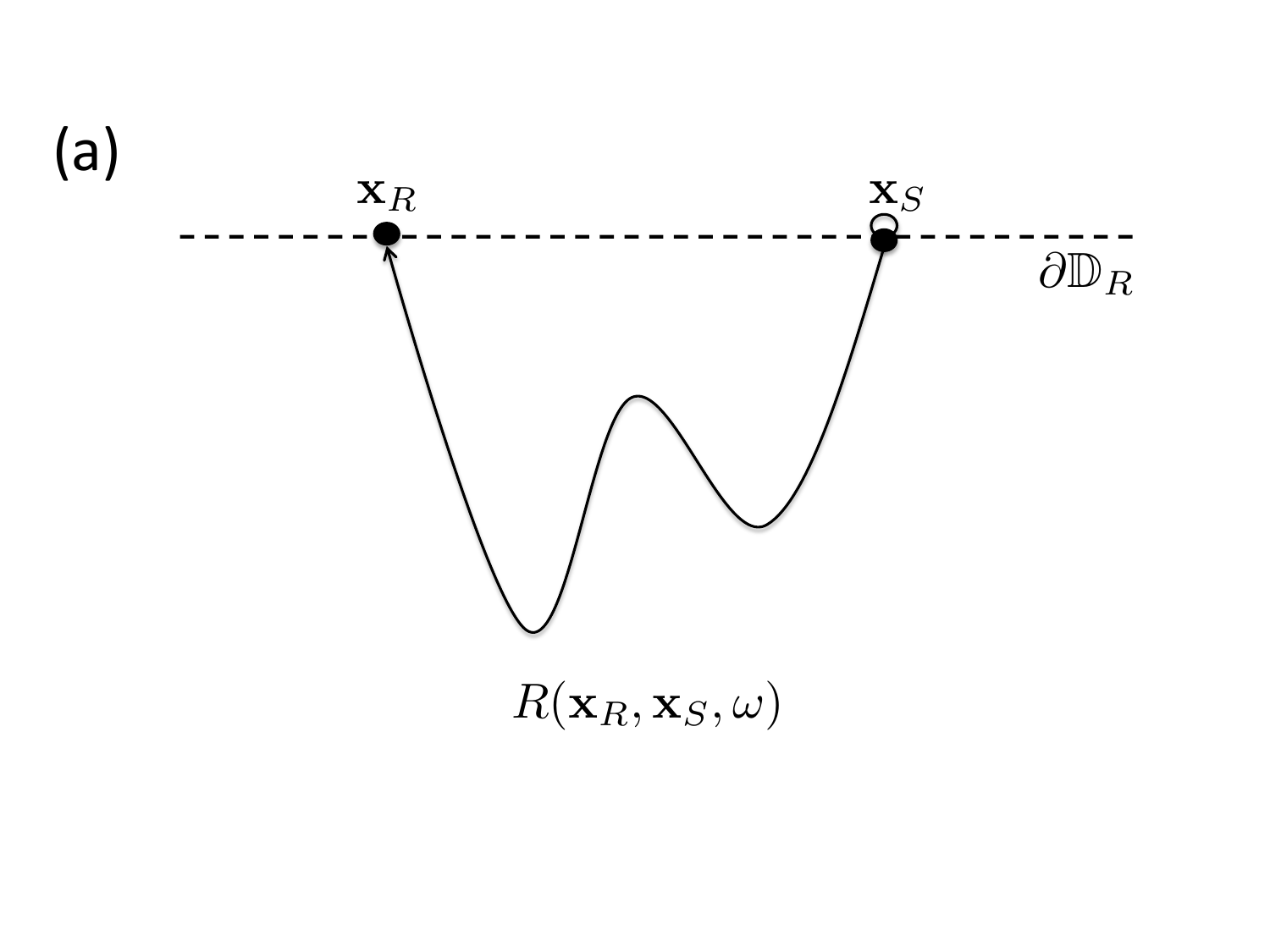}}
\vspace{-1.cm}
\centerline{\epsfysize=7 cm \epsfbox{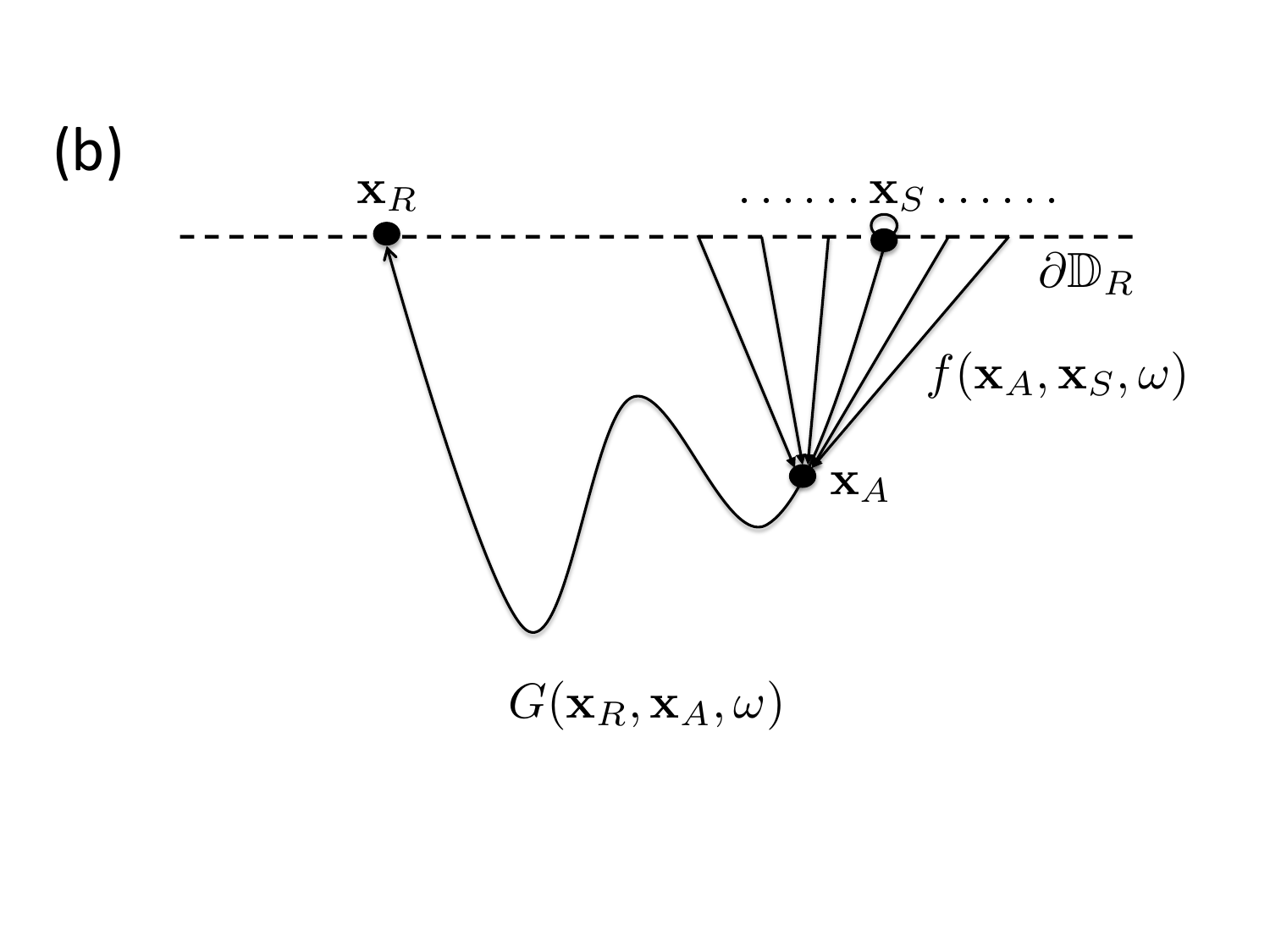}}
\vspace{-1.cm}
\centerline{\epsfysize=7 cm \epsfbox{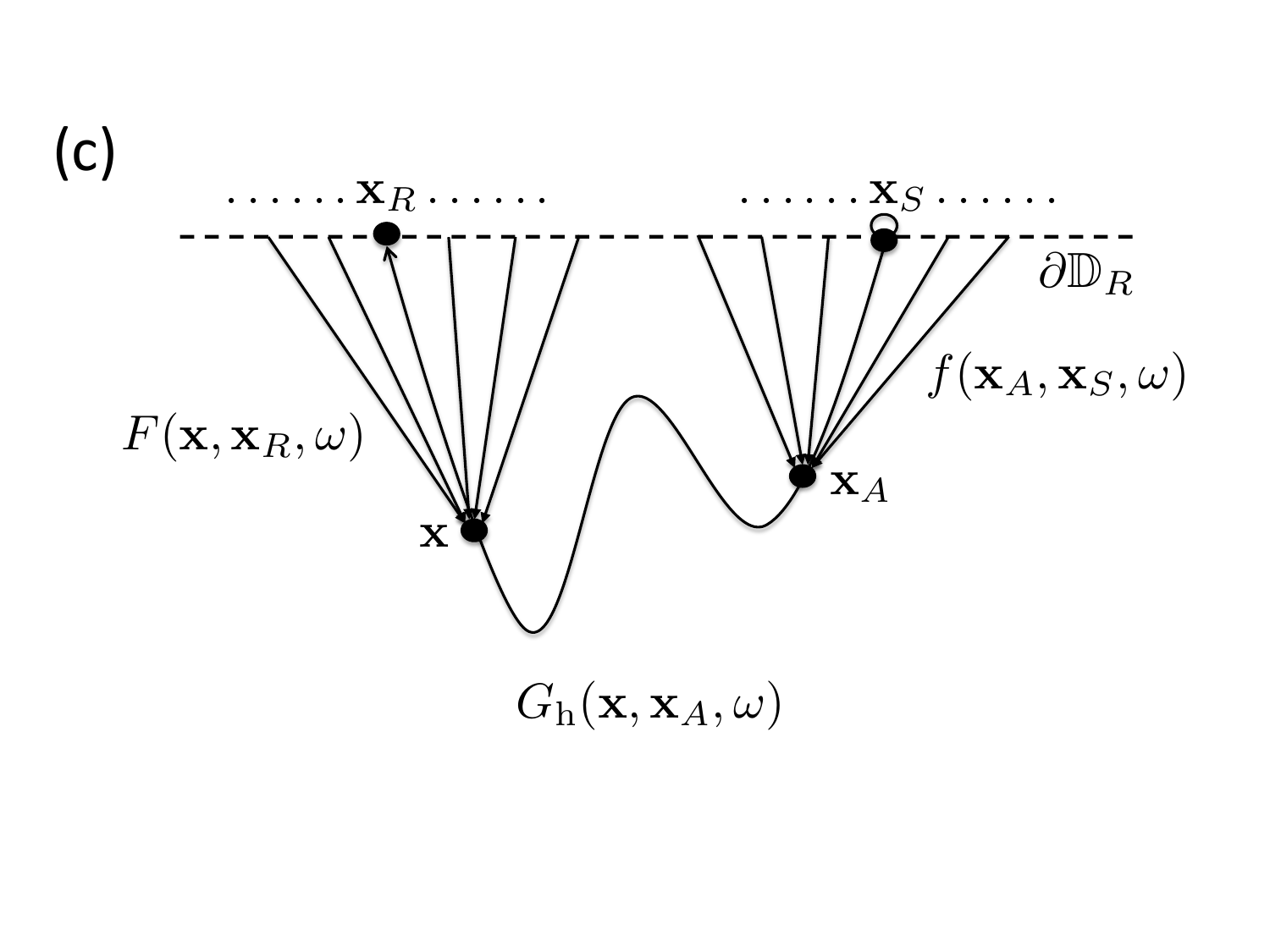}}
\vspace{-1.cm}
\caption{Illustration of source and receiver redatuming as a two-step process. 
Starting with (a) the reflection response $R({\bf x}_R,{\bf x}_S,\omega)$ at the surface, in step one (b) the Green's function $G({\bf x}_R,{{\bf x}_A},\omega)$ is obtained
for a virtual source at ${{\bf x}_A}$,
 and step two (c) yields the homogeneous Green's function $G_{\rm h}({\bf x},{{\bf x}_A},\omega)$ for a virtual receiver at ${\bf x}$. 
All functions in this figure are represented by simple rays, but in reality these are wave fields, including primaries, multiples, refracted and evanescent waves.
}\label{Figure2}
\end{figure}

We start by renaming the coordinate vectors in equation (\ref{eq56b}) as follows:
${\bf x}_S\to {\bf x}_R$,  ${\bf x}_R\to {\bf x}_S$, 
${\bf x}\to {{\bf x}_A}$. This yields, in combination with applying  source-receiver reciprocity on the left-hand side of equation (\ref{eq56b}),
\begin{eqnarray}
G({\bf x}_R,{{\bf x}_A},\omega)
&=&\int_{{{\partial\mathbb{D}}_R}} R({\bf x}_R,{\bf x}_S,\omega)f({{\bf x}_A},{\bf x}_S,\omega){\rm d}{\bf x}_S
+f^*({{\bf x}_A},{\bf x}_R,\omega),\,  \mbox{for}\quad x_{3,A} \ge x_{3,R}. \label{eq75}
\end{eqnarray}
Here $R({\bf x}_R,{\bf x}_S,\omega)$ is the reflection response  to a dipole source at ${\bf x}_S$, observed by 
a receiver at  ${\bf x}_R$, both at the surface ${{\partial\mathbb{D}}_R}$. This is schematically illustrated in Figure \ref{Figure2}(a).
The integral in  equation (\ref{eq75}) describes redatuming of the sources from all ${\bf x}_S$ at the surface to virtual-source position ${{\bf x}_A}$ in the subsurface, see Figure \ref{Figure2}(b).
After adding $f^*({{\bf x}_A},{\bf x}_R,\omega)$ (according to equation \ref{eq75})) this gives the Green's function $G({\bf x}_R,{{\bf x}_A},\omega)$, which is the response to the virtual monopole 
source at ${{\bf x}_A}$, observed by the receiver at ${\bf x}_R$ at the surface.

Our next aim is to derive a representation
for the response observed by a virtual receiver at ${\bf x}$ in the subsurface, given $G({\bf x}_R,{{\bf x}_A},\omega)$. Equation (\ref{eq12}) cannot be used for this in the same way as before, since
$G({\bf x},{{\bf x}_A},\omega)$ obeys a wave equation with a singularity at ${{\bf x}_A}$, whereas $p({\bf x},\omega)$ in equation (\ref{eq12}) is not allowed to have sources in the lower half-space.
To overcome this problem, we define the homogeneous Green's function \citep{Porter70JOSA, Oristaglio89IP}
\begin{eqnarray}
G_{\rm h}({\bf x},{{\bf x}_A},\omega)=G({\bf x},{{\bf x}_A},\omega)+G^*({\bf x},{{\bf x}_A},\omega).\label{eqGH}
\end{eqnarray}
Here $G({\bf x},{{\bf x}_A},\omega)$ and $G^*({\bf x},{{\bf x}_A},\omega)$ obey equation (\ref{eqweGom}),  with source terms 
$i\omega\delta({\bf x}-{{\bf x}_A})$ and $-i\omega\delta({\bf x}-{{\bf x}_A})$, respectively, on the right-hand sides.
Hence, $G_{\rm h}({\bf x},{{\bf x}_A},\omega)$ obeys the following equation
\begin{eqnarray}
{\cal L} G_{\rm h}=0,\label{eqweGhom}
\end{eqnarray}
which confirms that the homogeneous Green's function is source-free.
This time we choose for $p({\bf x},\omega)$ in equation (\ref{eq12})
\begin{eqnarray}
p({\bf x},\omega)&=&G_{\rm h}({\bf x},{{\bf x}_A},\omega),\label{eq31H}
\end{eqnarray}
with $G_{\rm h}({\bf x},{{\bf x}_A},\omega)$ defined in equation (\ref{eqGH}).
For ${\bf x}$ at ${{\partial\mathbb{D}}_R}$ the Green's function $G({\bf x},{{\bf x}_A},\omega)$ is purely upgoing, since the upper half-space is homogeneous and 
the virtual source at ${{\bf x}_A}$ lies in the lower half-space.
Similarly, $G^*({\bf x},{{\bf x}_A},\omega)$ is downgoing at ${{\partial\mathbb{D}}_R}$, except for the evanescent field (which we already neglected at ${{\partial\mathbb{D}}_R}$ in the derivation of equation (\ref{eq12})). 
Hence, we may write
\begin{eqnarray}
p^-({\bf x},\omega)&=&G_{\rm h}^-({\bf x},{{\bf x}_A},\omega)=  G({\bf x},{{\bf x}_A},\omega),\quad\mbox{for}\quad x_3=x_{3,R},\label{eq33HH}\\
p^+({\bf x},\omega)&=&G_{\rm h}^+({\bf x},{{\bf x}_A},\omega)=  G^*({\bf x},{{\bf x}_A},\omega), \quad\mbox{for}\quad x_3=x_{3,R}.\label{eq33H}
\end{eqnarray}
Substitution of equations (\ref{eq31H}) $-$ (\ref{eq33H}) into equation (\ref{eq12}) yields
\begin{eqnarray}
G_{\rm h}({\bf x},{{\bf x}_A},\omega)&=&\int_{{{\partial\mathbb{D}}_R}} F({\bf x},{\bf x}_R,\omega)G({\bf x}_R,{{\bf x}_A},\omega){\rm d}{\bf x}_R 
+\int_{{{\partial\mathbb{D}}_R}} F^*({\bf x},{\bf x}_R,\omega)G^*({\bf x}_R,{{\bf x}_A},\omega){\rm d}{\bf x}_R, \nonumber\\
&&\hspace{8cm}\mbox{for}\quad x_3\ge x_{3,R} , \label{eq12abx}
\end{eqnarray}
or
\begin{eqnarray}
G_{\rm h}({\bf x},{{\bf x}_A},\omega)&=&2\Re\int_{{{\partial\mathbb{D}}_R}} F({\bf x},{\bf x}_R,\omega)G({\bf x}_R,{{\bf x}_A},\omega){\rm d}{\bf x}_R, 
\quad\mbox{for}\quad x_3\ge x_{3,R}, \label{eq12ab}
\end{eqnarray}
where $\Re$ denotes that the real part is taken.
For an intuitive explanation of the right-hand side of equation (\ref{eq12abx}) we refer again to Figure \ref{FigA2}, this time with $p^+({\bf x}_R,\omega)$ and $p^-({\bf x}_R,\omega)$ replaced
by $G^*({\bf x}_R,{{\bf x}_A},\omega)$ and $G({\bf x}_R,{{\bf x}_A},\omega)$, respectively. The focusing functions propagate these downgoing and upgoing Green's functions at ${\bf x}_R$ from
${{{\partial\mathbb{D}}_R}}$ into the lower half-space, with the focal points acting as secondary sources in a modified form of Huygens' principle. 
The two integrals  cannot be separately associated with $G^*({\bf x},{{\bf x}_A},\omega)$ and $G({\bf x},{{\bf x}_A},\omega)$ for ${\bf x}$
 in the lower half-space (these functions are singular at ${{\bf x}_A}$);
only the sum of the two integrals gives $G_{\rm h}({\bf x},{{\bf x}_A},\omega)$ (which is not singular at ${{\bf x}_A}$).

Hence, equation (\ref{eq12abx}) describes redatuming of the receivers from all ${\bf x}_R$ at the surface to virtual-receiver position ${\bf x}$ in the subsurface, see Figure \ref{Figure2}(c).
It gives the homogeneous Green's function $G_{\rm h}({\bf x},{{\bf x}_A},\omega)$, which is the response to the virtual source at ${{\bf x}_A}$, observed by
a virtual receiver at ${\bf x}$, plus its complex conjugate. After transforming this to the time domain we obtain
\begin{eqnarray}
G_{\rm h}({\bf x},{{\bf x}_A},t)=G({\bf x},{{\bf x}_A},t)+G({\bf x},{{\bf x}_A},-t).
\end{eqnarray}
The two functions at the right-hand side of this equation do not overlap in time (except for ${\bf x}={{\bf x}_A}$ and only for $t=0$), 
hence, $G({\bf x},{{\bf x}_A},t)$ can be extracted from $G_{\rm h}({\bf x},{{\bf x}_A},t)$ by selecting 
its causal part.

Note that there is an asymmetry in the focusing functions used for source redatuming ($f({{\bf x}_A},{\bf x}_S,\omega)$ in equation (\ref{eq75})) and for receiver redatuming
($F({\bf x},{\bf x}_R,\omega)$ in equation (\ref{eq12ab})), see also Figure \ref{Figure2}(c). 
This is due to the difference in types of responses at the surface (the dipole response $R({\bf x}_R,{\bf x}_S,\omega)$) 
and in the subsurface (the monopole response $G({\bf x},{{\bf x}_A},\omega)$). When the response at the surface were also a monopole response, then the 
focusing function $f({{\bf x}_A},{\bf x}_S,\omega)$ for source redatuming should be replaced by $F({{\bf x}_A},{\bf x}_S,\omega)$.

Homogeneous Green's function representations similar to equation (\ref{eq12ab}) were also derived by \citet{Wapenaar2016GJI}, \citet{Neut2017JASA} and \citet{Singh2017GEO2}, 
but here equation (\ref{eq12ab}) has been derived without up/down decomposition inside the medium. 
Hence, it also holds for evanescent waves inside the medium, as long as condition (\ref{eqevan}) is obeyed.
 Moreover, the derivation presented here is much simpler than in those references.

The source and receiver redatuming processes can be captured in one equation by substituting equation (\ref{eq75}) into  (\ref{eq12ab}). This gives
\begin{eqnarray}
G_{\rm h}({\bf x},{{\bf x}_A},\omega)&=&
2\Re\int_{{{\partial\mathbb{D}}_R}}\int_{{{\partial\mathbb{D}}_R}} F({\bf x},{\bf x}_R,\omega)R({\bf x}_R,{\bf x}_S,\omega)f({{\bf x}_A},{\bf x}_S,\omega){\rm d}{\bf x}_S{\rm d}{\bf x}_R \nonumber\\
&+&2\Re\int_{{{\partial\mathbb{D}}_R}} F({\bf x},{\bf x}_R,\omega)f^*({{\bf x}_A},{\bf x}_R,\omega){\rm d}{\bf x}_R,
\quad\mbox{for}\quad \{x_3,x_{3,A}\}\ge x_{3,R}. \label{eq12abk}
\end{eqnarray}
The double integral on the right-hand side resembles the process of classical source and receiver redatuming \citep{Berkhout82Book,Berryhill84GEO}, but with
the primary focusing functions in those references replaced by full-field focusing functions. It also resembles source-receiver interferometry \citep{Curtis2010PRE}, but with the
double integration along a closed boundary in that paper replaced by the double integration over the open boundary ${{\partial\mathbb{D}}_R}$.
Hence, via the theories of primary source-receiver redatuming \citep{Berkhout82Book,Berryhill84GEO}, closed-boundary source-receiver interferometry \citep{Curtis2010PRE}
and open-boundary homogeneous Green's function retrieval using wave field decomposition \citep{Wapenaar2016GJI, Neut2017JASA, Singh2017GEO2}, 
we have arrived at a representation for open-boundary homogeneous full-field Green's function retrieval  (equation \ref{eq12abk}), 
which accounts for internal multiples, and refracted and evanescent waves in the lower half-space. In section \ref{sec5.3} this representation is extended for the elastodynamic situation.

\subsection{Numerical examples}\label{sec3.4}

We illustrate the representations of sections \ref{sec3.2} and \ref{sec3.3}
with numerical examples. Our main aim  is to demonstrate that these representations hold for evanescent waves
inside the medium.
To this end we consider oblique plane waves in a horizontally layered medium, 
with isotropic depth-dependent medium parameters $c(x_3)$ (propagation velocity) and $\rho(x_3)$ (mass density).
We consider a horizontally layered medium because in this case we can unequivocally distinguish between propagating and evanescent waves. However,
as discussed in section \ref{sec2}, the representations also account for  evanescent waves in more general inhomogeneous media.
We define the spatial Fourier transform of a space- and frequency-dependent function $u({\bf x},\omega)$ as 
\begin{eqnarray}
&&\hspace{-0.5cm}\tilde u({\bf s},x_3,\omega)=\int_{\mathbb{R}^2}\exp\{-i\omega{\bf s}\cdot{\bf x}_{\rm H}\}u({\bf x}_{\rm H},x_3,\omega){\rm d}{\bf x}_{\rm H},\label{eq50a}
\end{eqnarray}
with ${\bf s}=(s_1,s_2)$, where $s_1$ and $s_2$ are horizontal slownesses and $\mathbb{R}$ is the set of real numbers. This decomposes the function $u({\bf x},\omega)$ into monochromatic 
plane-wave components. Next, we define the inverse temporal Fourier transform per slowness value as
\begin{eqnarray}
u({\bf s},x_3,\tau)=\frac{1}{\pi}\Re\int_0^\infty \tilde u({\bf s},x_3,\omega)\exp\{-i\omega\tau){\rm d}\omega,\label{eq500a}
\end{eqnarray}
where $\tau$ is the so-called intercept time \citep{Stoffa89Book}.

\begin{figure}
\vspace{-.0cm}
\centerline{\epsfysize=7. cm \epsfbox{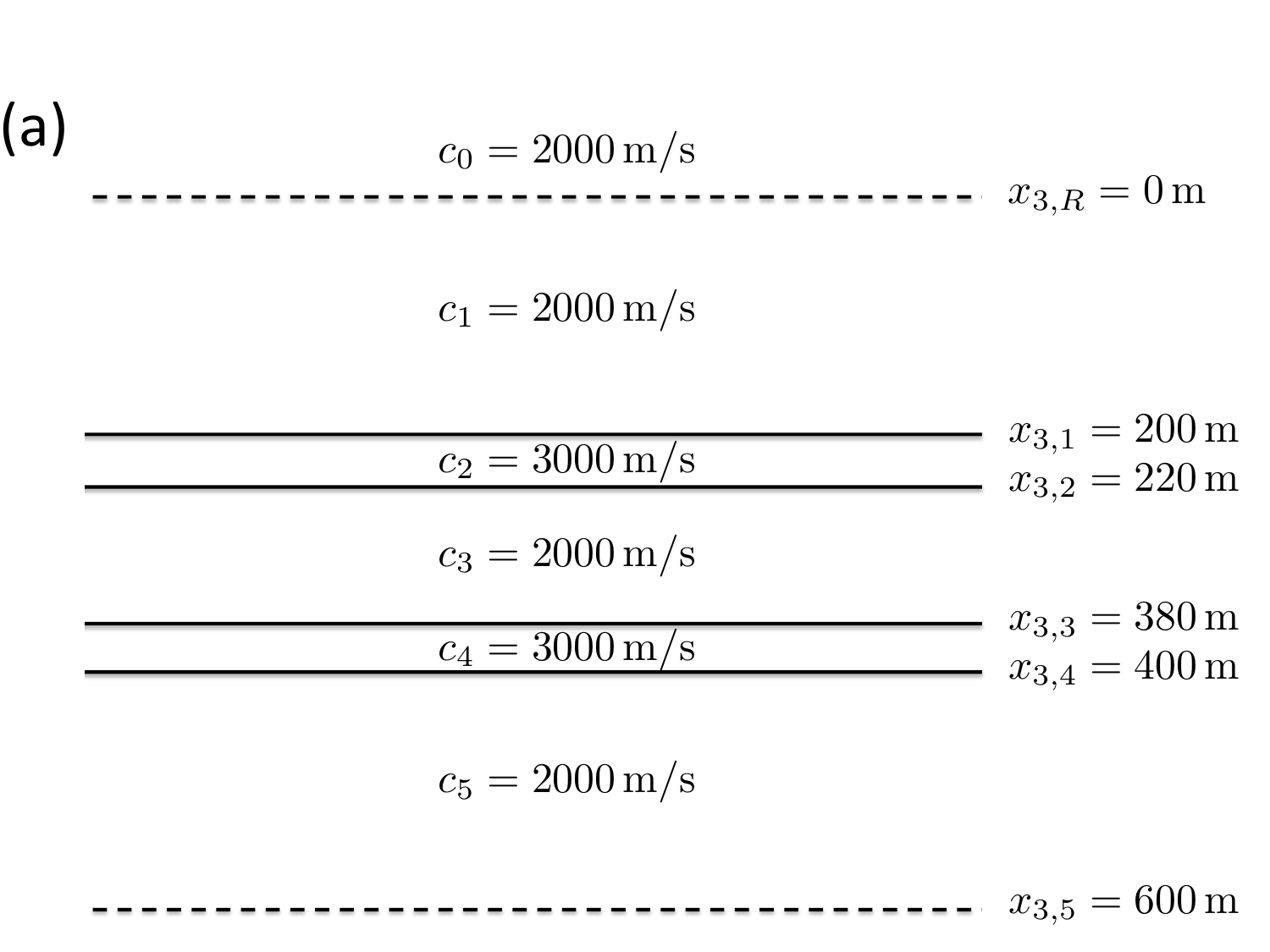}}
\vspace{.5cm}
\centerline{\epsfysize=7. cm \epsfbox{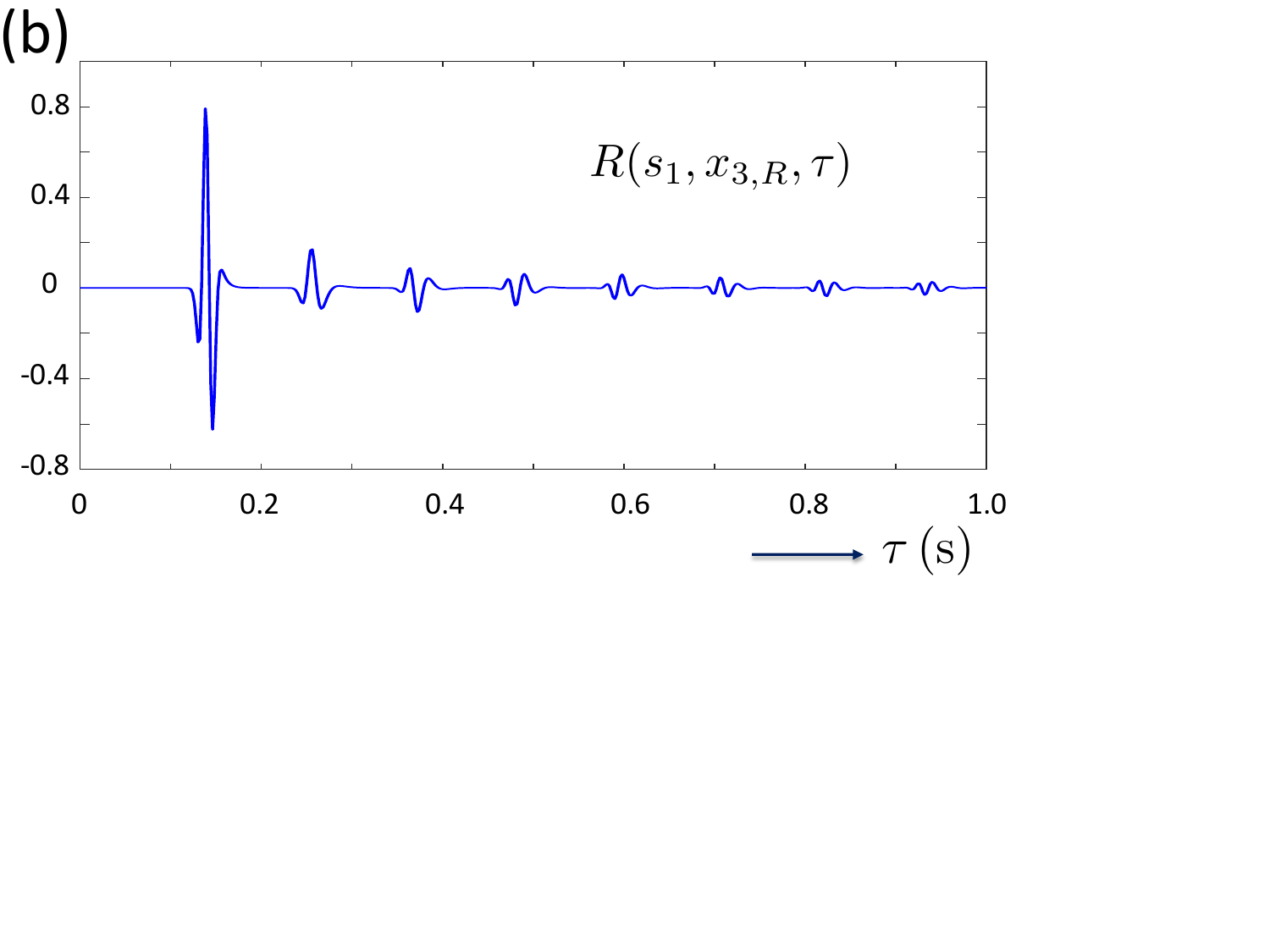}}
\vspace{-1.5cm}
\centerline{\epsfysize=7. cm \epsfbox{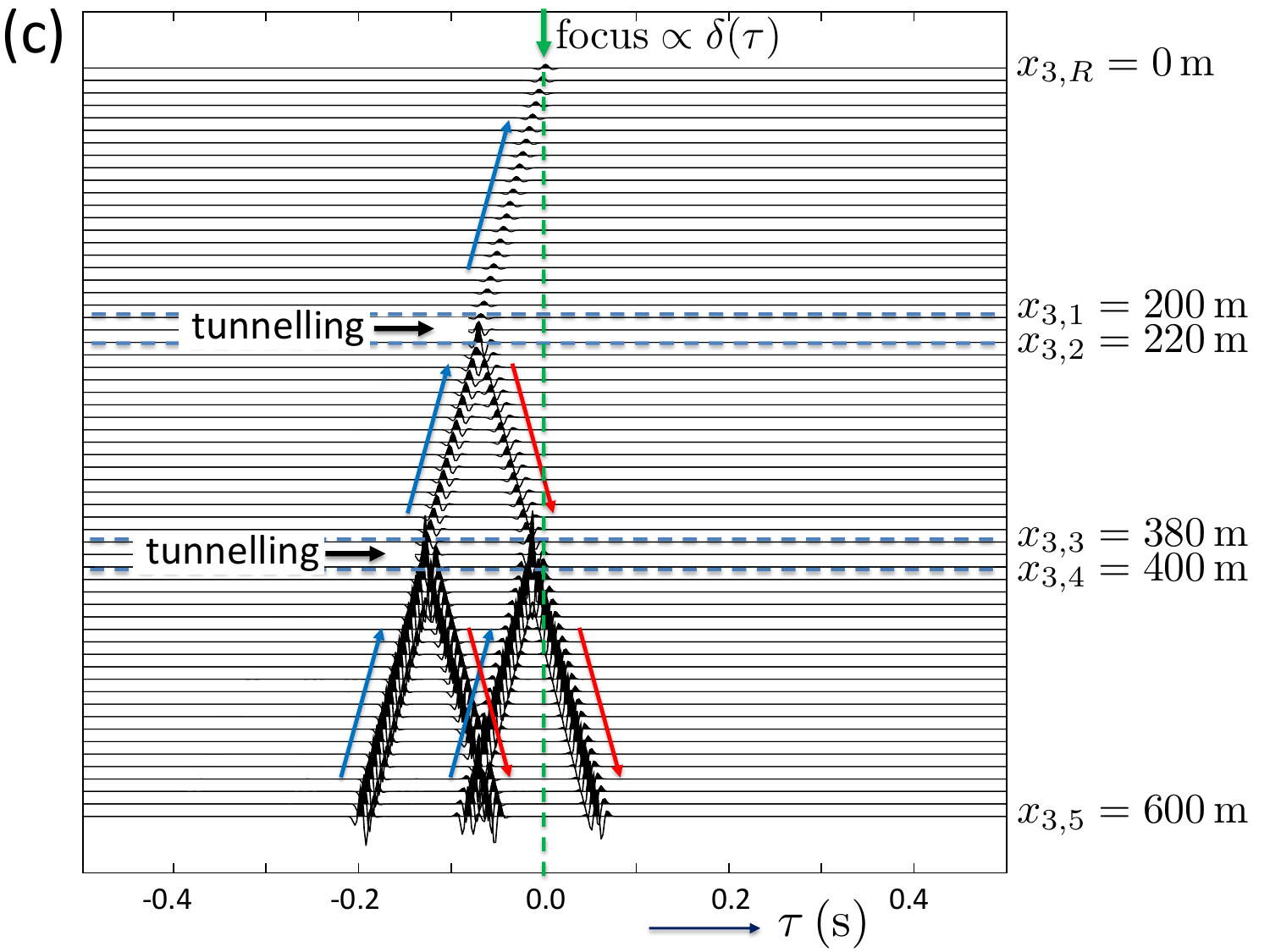}}
\vspace{-.cm}
\caption{(a) Horizontally layered medium with two high-velocity layers. (b) Numerically modelled reflection response $R(s_1,x_{3,R},\tau)$ at the surface. The horizontal slowness $s_1=1/2800$ s m$^{-1}$
is chosen such that the wave field is evanescent in the high-velocity layers. (c) Numerically modelled focusing function $f(s_1,x_3,x_{3,R},\tau)$. The trace at $x_{3,R}=0$ m illustrates the focusing condition
of equation (\ref{eqfocp}).
}\label{Figure3}
\end{figure}

First we investigate the representation of equation (\ref{eq56b}) and take ${\bf x}_S=(0,0,x_{3,R})$. 
We use the definitions of equations (\ref{eq50a}) and (\ref{eq500a}) to transform this representation to the slowness intercept-time domain.
Taking into account that for a horizontally layered, isotropic medium all functions in  equation (\ref{eq56b}) are cylindrically symmetric, 
it suffices to consider the transformed representation for one slowness variable  only. We thus obtain 
\begin{eqnarray}
G(s_1,x_3,x_{3,R},\tau)
&=&\int_{-\infty}^\tau f(s_1,x_3,x_{3,R},\tau')R(s_1,x_{3,R},\tau-\tau'){\rm d}\tau'\nonumber\\
&+&f(s_1,x_3,x_{3,R},-\tau),\quad\mbox{for}\quad x_3\ge x_{3,R}.\label{eq20}
\end{eqnarray}
For any given value of $s_1$, the Green's function $G(s_1,x_3,x_{3,R},\tau)$ is the response to a plane-wave source at $x_{3,R}$ as a function of receiver depth $x_3$ and intercept time $\tau$. 
For $|s_1|\le 1/c(x_3)$ the plane wave is propagating, whereas for $|s_1|> 1/c(x_3)$ it is evanescent. 
For propagating waves, the local propagation angle $\alpha(x_3)$ follows from $s_1=\sin\alpha(x_3)/c(x_3)$.
The focusing function $f(s_1,x_3,x_{3,R},\tau)$ obeys the 
focusing condition formulated by equation (\ref{eqfoc2b}), transformed to the slowness intercept-time domain, hence
\begin{eqnarray}
f(s_1,x_3,x_{3,R},\tau)|_{x_3=x_{3,R}}&=&\frac{\rho_0c_0}{2\cos\alpha_0}\delta(\tau),\label{eqfocp}
\end{eqnarray}
with $\alpha_0=\alpha(x_{3,R})$. Consider the horizontally layered medium of Figure \ref{Figure3}(a). Two thin high-velocity layers ($c_2=c_4=3000$ m/s) are embedded
in a homogeneous background medium with a velocity of 2000 m/s. The mass densities, in kg m$^{-3}$, 
are assigned the same numerical values as the velocities to get significant contrasts between the different layers.
A plane wave is emitted from $x_{3,R}$ into the medium, with slowness $s_1=1/2800$ s m$^{-1}$, hence, this wave leaves the surface with an angle $\alpha_0=45.6^o$ and becomes evanescent in the
high-velocity layers. For the source function we use a Ricker wavelet with a central frequency of 50 Hz, hence, the wavelength for the central frequency in the high-velocity layers is 60 m. The thickness 
of the high-velocity layers is 20 m, which is of the same order as the distance over which the evanescent waves decay with a factor 1/e, which is equal to 
$1/(\omega_c\sqrt{s_1^2-1/c_2^2})$=24.8 m. Hence, we may expect that the waves tunnel through these layers. Figure \ref{Figure3}(b) shows the numerically modelled
reflection response $R(s_1,x_{3,R},\tau)$ for the chosen slowness. The first two events are composite reflections 
from the two high-velocity layers (including internal multiples of evanescent waves inside these layers) 
and the other events are multiple reflections between these layers.
Figure \ref{Figure3}(c) shows the numerically modelled focusing function $f(s_1,x_3,x_{3,R},\tau)$ as a function of $x_3$ and $\tau$, convolved with the 
same Ricker wavelet for a clear display. Blue and red arrows indicate upgoing and downgoing waves, respectively,
in the homogeneous background medium. The tunnelling of the waves through the high-velocity layers is  clearly visible. A single upgoing 
wave reaches the surface $x_{3,R}$ at $\tau=0$, conform the focusing condition formulated by equation (\ref{eqfocp}) (except that in this display $\delta(\tau)$ is convolved with the Ricker wavelet).
Note that the amplitude increases with increasing depth (to compensate for the evanescent waves in the high-velocity layers), which means that, in practice, 
 the numerically computed focusing function becomes unstable beyond some thickness of the high-velocity layers.

\begin{figure}
\vspace{-.0cm}
\centerline{\epsfysize=6. cm \epsfbox{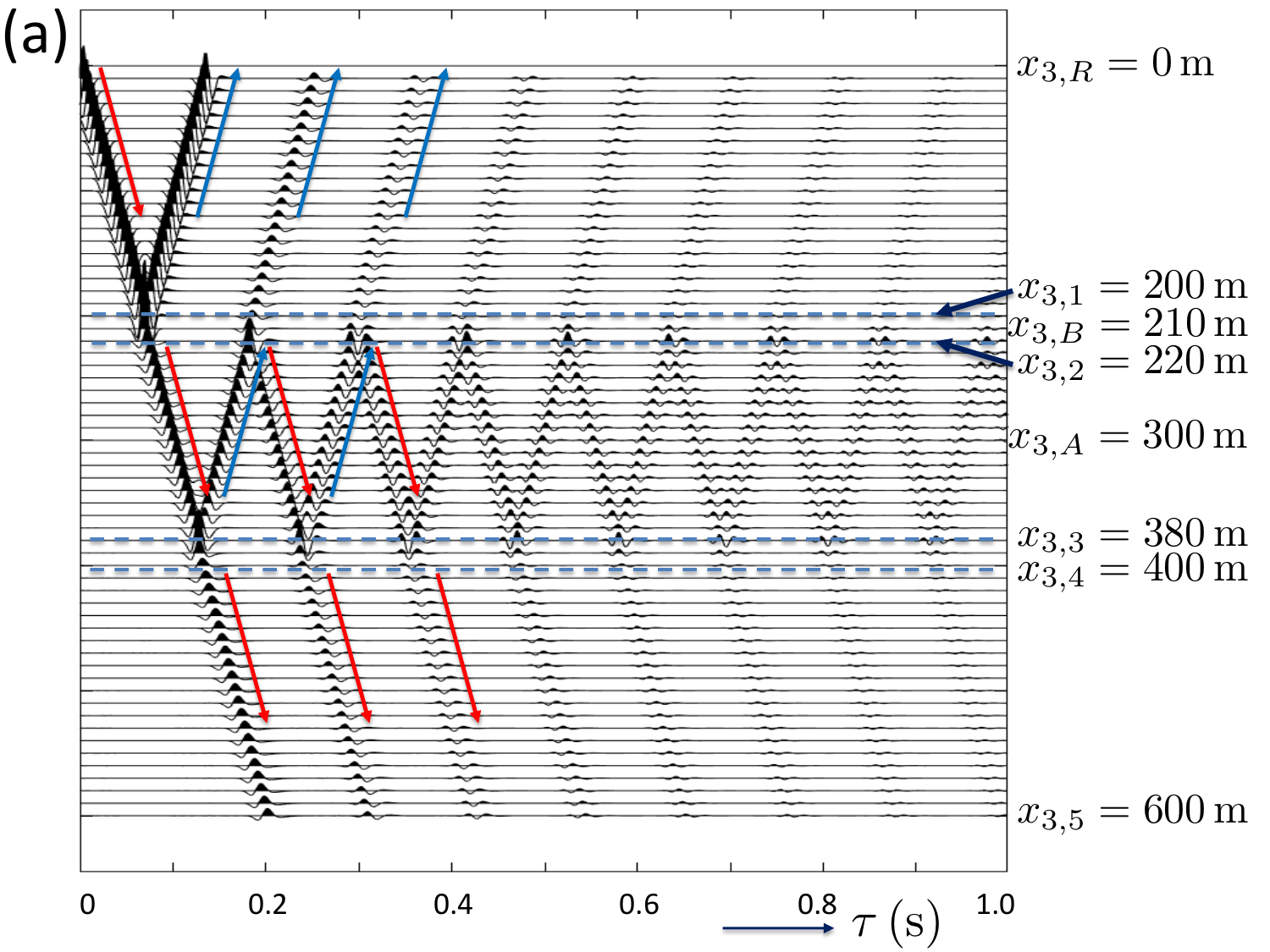}}
\vspace{.5cm}
\centerline{\epsfysize=6. cm \epsfbox{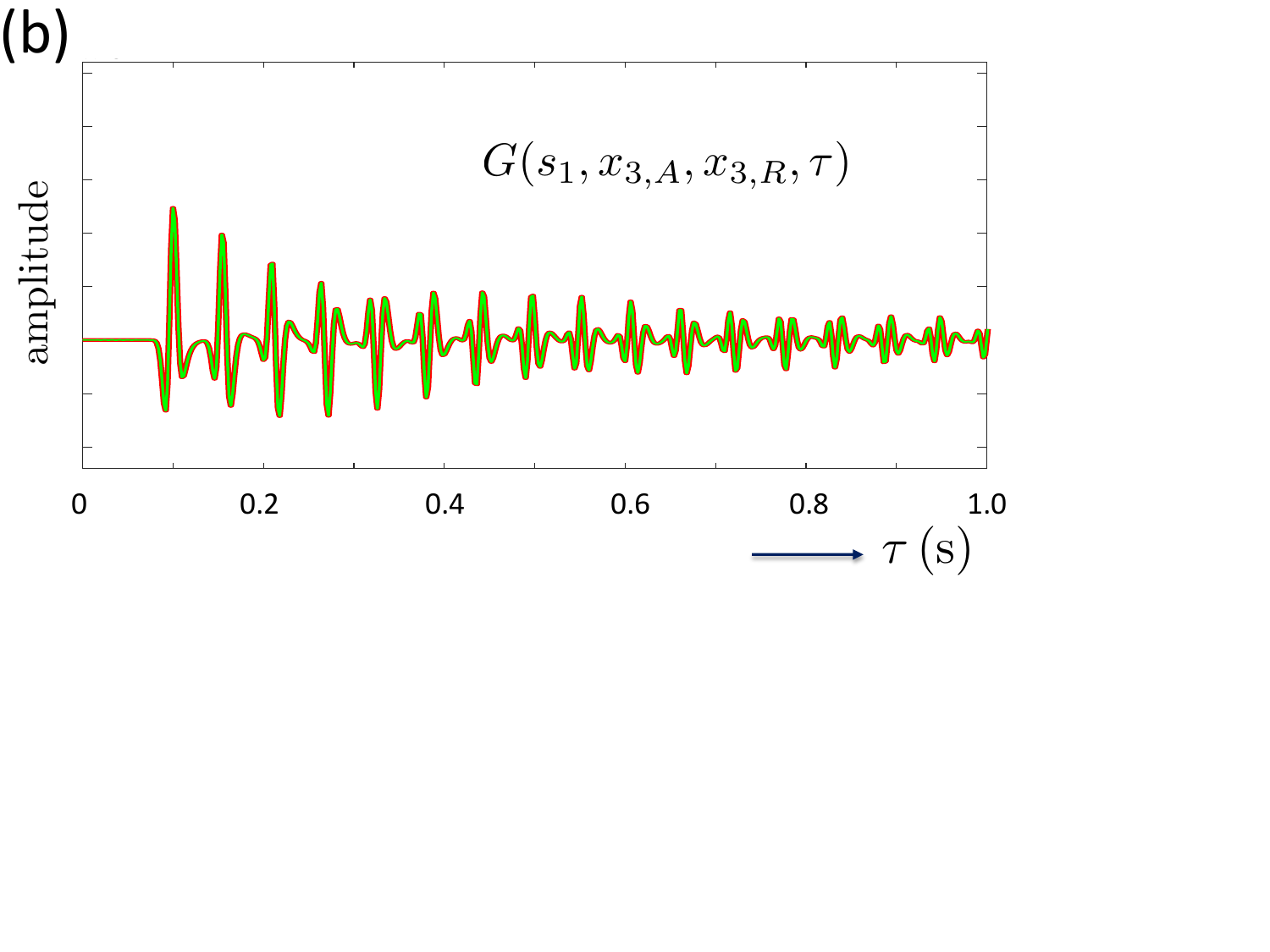}}
\vspace{-2.cm}
\centerline{\epsfysize=6. cm \epsfbox{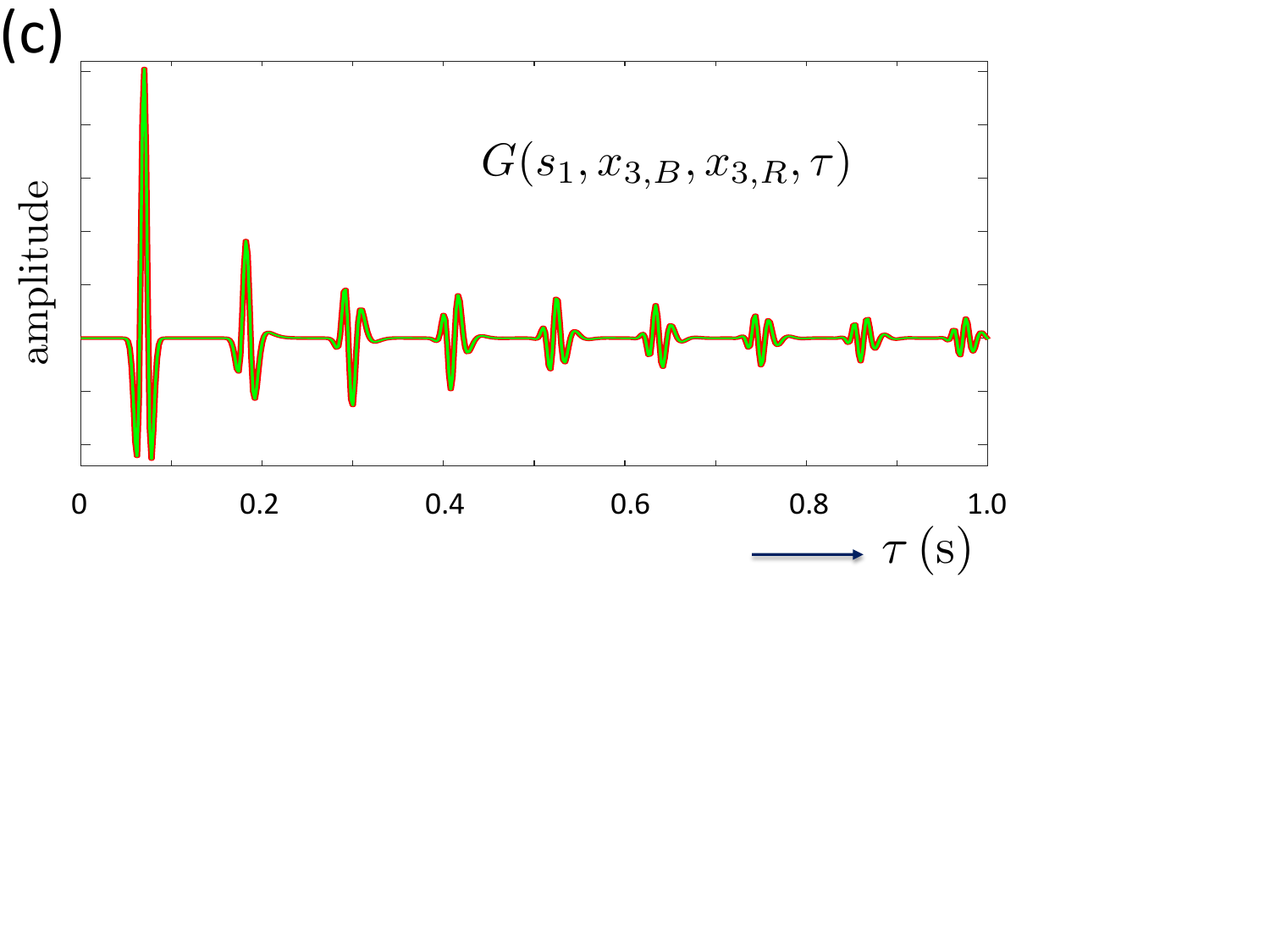}}
\vspace{-2cm}
\centerline{\epsfysize=6. cm \epsfbox{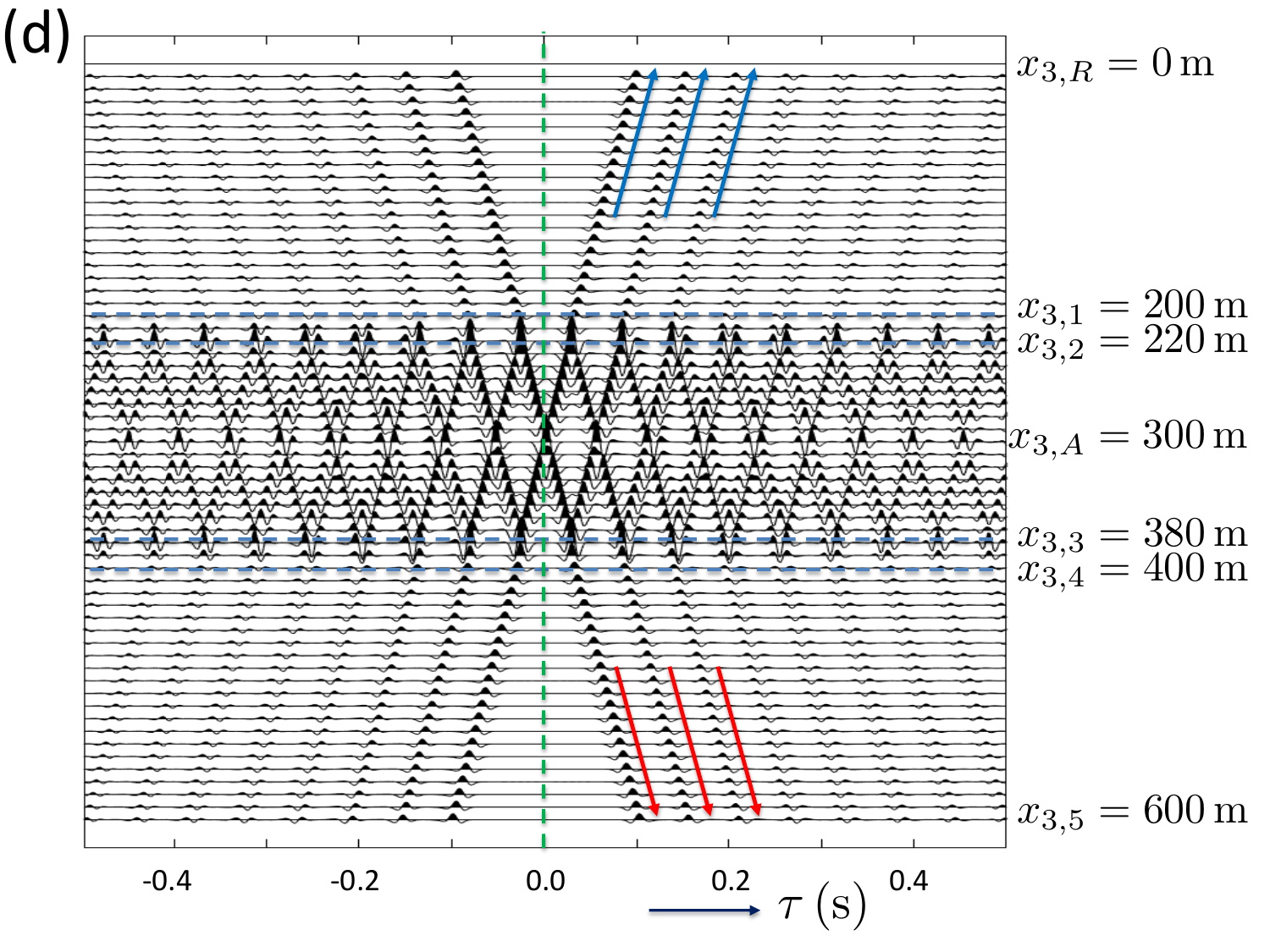}}
\vspace{-.cm}
\caption{(a) Green's function $G(s_1,x_3,x_{3,R},\tau)$ obtained from Figures \ref{Figure3}(b) and \ref{Figure3}(c) via the representation of equation (\ref{eq20}).
(b) $G(s_1,x_{3,A},x_{3,R},\tau)$, taken from figure (a) for $x_{3,A}=300$ m (green), compared with directly modelled Green's function (red).
(c) Similarly,  $G(s_1,x_{3,B},x_{3,R},\tau)$, taken from figure (a) for $x_{3,B}=210$ m inside the first high velocity layer. 
(d) Homogeneous Green's function $G_{\rm h}(s_1,x_3,x_{3,A},\tau)$ obtained from Figures \ref{Figure3}(c) and \ref{Figure4}(b) via the representation of equation (\ref{eq20p}).
}\label{Figure4}
\end{figure}

The reflection response of Figure \ref{Figure3}(b) and the focusing function of Figure \ref{Figure3}(c) (the latter without the wavelet) are used as input for the representation of equation (\ref{eq20}).
 This yields the Green's function $G(s_1,x_3,x_{3,R},\tau)$ (convolved with the Ricker wavelet) as a function of $x_3$ and $\tau$, see Figure \ref{Figure4}(a). 
 Blue and red arrows  indicate again upgoing and downgoing waves, respectively.
 This figure shows the expected behaviour of the response to a plane-wave source at $x_{3,R}$
 (a downgoing wave leaving the surface, two composite primary upgoing waves and  multiple reflections between the high velocity layers). Figure \ref{Figure4}(b) shows
 $G(s_1,x_{3,A},x_{3,R},\tau)$ for $x_{3,A}=300$ m. The green line is the Green's function obtained from equation (\ref{eq20}), the red line is the directly modelled 
 Green's function. Similarly,  Figure \ref{Figure4}(c) shows $G(s_1,x_{3,B},x_{3,R},\tau)$ for $x_{3,B}=210$ m, i.e., inside the first high velocity layer. 
 In both cases the match is perfect, which confirms that the representation of equation (\ref{eq20}) correctly accounts for propagating and evanescent waves inside the medium.
 
Using source-receiver reciprocity we may interpret Figure \ref{Figure4}(b) as $G(s_1,x_{3,R},x_{3,A},\tau)$, which is the response at the surface
$x_{3,R}$ to a virtual plane-wave source at $x_{3,A}=300$ m. Hence, $G(s_1,x_{3,R},x_{3,A},\tau)$ may be seen as the result of redatuming the source from the surface to $x_{3,A}$. 
We now discuss receiver redatuming.
To this end, we transform the representation of equation (\ref{eq12abx}) to the slowness intercept-time domain, which yields
\begin{eqnarray}
G_{\rm h}(s_1,x_3,x_{3,A},\tau)
&=&\int_{-\infty}^\tau F(s_1,x_3,x_{3,R},\tau')G(s_1,x_{3,R},x_{3,A},\tau-\tau'){\rm d}\tau'\nonumber\\
&+&\int_\tau^\infty F(s_1,x_3,x_{3,R},-\tau')G(s_1,x_{3,R},x_{3,A},\tau'-\tau){\rm d}\tau',\nonumber\\
&&\hspace{5cm}\mbox{for}\quad x_3\ge x_{3,R},\label{eq20p}
\end{eqnarray}
with, analogous to equation (\ref{eq71}),
\begin{eqnarray}
F(s_1,x_3,x_{3,R},\tau)=\frac{2\cos\alpha_0}{\rho_0c_0}f(s_1,x_3,x_{3,R},\tau).
\end{eqnarray}
Note that the right-hand side of equation (\ref{eq20p}) contains the  Green's function with the redatumed source at  $x_{3,A}$ and the receiver at  $x_{3,R}$  at the surface. 
This representation redatums the receiver from $x_{3,R}$ to any depth $x_3$ in the subsurface.
This yields the homogeneous Green's function, which consists of $G(s_1,x_3,x_{3,A},\tau)$ plus its time-reversal, see Figure \ref{Figure4}(d). The causal part (right of the green dashed line) is the retrieved
Green's function $G(s_1,x_3,x_{3,A},\tau)$. Conform expectation, 
we see a virtual source at $x_{3,A}$ emitting downgoing and upgoing plane waves, which reverberate in the wave guide between the two high-velocity
layers, but which also emit some energy through tunnelling into the half-spaces above and below the high-velocity layers. 
This example illustrates the handling of propagating and evanescent waves inside the medium by the homogeneous Green's function representation of equation (\ref{eq20p}).

\section{Elastodynamic wave field representation}

We derive the elastodynamic equivalent of the representation of equation (\ref{eq12}). We consider the same configuration as in section \ref{sec2}, except that now the  medium parameters
of the lower half-space $x_3>x_{3,R}$
are the stiffness tensor $c_{ijkl}({\bf x})$  and the mass density tensor $\rho_{ik}({\bf x})$, with symmetries  $c_{ijkl}=c_{jikl}=c_{ijlk}=c_{klij}$ and  $\rho_{ik}=\rho_{ki}$. 
In the homogeneous isotropic upper half-space $x_3\le x_{3,R}$ the parameters are $\rho_{ik}=\delta_{ik}\rho_0$ and $c_{ijkl}=\lambda_0\delta_{ij}\delta_{kl}+\mu_0(\delta_{ik}\delta_{jl}+\delta_{il}\delta_{jk})$,
with $\lambda_0$ and $\mu_0$ the Lam\'e parameters of the half-space. The $P$- and $S$-wave propagation velocities of the upper half-space are $c_P=((\lambda_0+2\mu_0)/\rho_0)^{1/2}$
and $c_S=(\mu_0/\rho_0)^{1/2}$, respectively.

The basic equations in the frequency domain for elastodynamic wave propagation are the linearized equation of motion
\begin{eqnarray}
-i\omega\rho_{ik}v_k-\partial_j\tau_{ij}&=&\hat f_i\label{eq433}
\end{eqnarray}
and the linearized deformation equation
\begin{eqnarray}
i\omega\tau_{ij}+c_{ijkl}\partial_lv_k&=&0,\label{eq434}
\end{eqnarray}
respectively. Here $\tau_{ij}({\bf x},\omega)$ is the stress tensor (with symmetry $\tau_{ij}=\tau_{ji}$), $v_k({\bf x},\omega)$ the particle velocity and $\hat f_i({\bf x},\omega)$ a source in terms of volume-force density 
(the circumflex is used to distinguish this source term from the focusing function). Equations (\ref{eq433}) and (\ref{eq434}) can be combined into the elastodynamic wave equation
\begin{eqnarray}
{\cal L}_{ik}v_k=i\omega\hat f_i,\label{eqweedyn}
\end{eqnarray}
with
\begin{eqnarray}
{\cal L}_{ik}=\partial_jc_{ijkl}\partial_l+\omega^2\rho_{ik}.
\end{eqnarray}
We introduce an elastodynamic focusing function ${\bf F}({\bf x},{\bf x}_R,\omega)$ as a $3\times 3$ matrix, according to
\begin{eqnarray}
{\bf F}({\bf x},{\bf x}_R,\omega)=
\begin{pmatrix}
F_{1,1} & F_{1,2} & F_{1,3} \\
F_{2,1} & F_{2,2} & F_{2,3} \\
F_{3,1} & F_{3,2} & F_{3,3} 
\end{pmatrix}({\bf x},{\bf x}_R,\omega),\label{eq301}
\end{eqnarray}
where ${\bf x}_R$ denotes again the position of a focal point at ${{\partial\mathbb{D}}_R}$.
Each column of ${\bf F}$ is a particle velocity vector of which the components, for fixed ${\bf x}_R$ and variable ${\bf x}$, obey the elastodynamic wave equation (\ref{eqweedyn}) for the source-free situation.
This is different from the elastodynamic focusing function introduced by \cite{Wapenaar2014GJI}, in which the different elements represent decomposed compressional and shear waves.

We define the  focusing condition, analogous to equation (\ref{eq4}), as
\begin{eqnarray}
{\bf F}({\bf x},{\bf x}_R,\omega)|_{x_3=x_{3,R}}&=&{\bf I}\delta({\bf x}_{\rm H}-{\bf x}_{{\rm H},R}),\label{eq302}
\end{eqnarray}
(${\bf I}$ is the $3\times 3$ identity matrix) and demand that ${\bf F}({\bf x},{\bf x}_R,\omega)$ is purely upgoing at ${{\partial\mathbb{D}}_R}$ and  in the homogeneous isotropic upper half-space.
Equation (\ref{eq302}) implies that, for the $k$th column of ${\bf F}$, the $k$th component of the particle velocity vector in that column focuses at ${\bf x}_R$ 
and the other two components  are zero on ${{\partial\mathbb{D}}_R}$. 
Hence, the columns of ${\bf F}$ are mutually independent.

We discuss a representation for a wave field $v_k({\bf x},\omega)$, which may have sources in the upper half-space above ${{\partial\mathbb{D}}_R}$, 
but which obeys the source-free wave equation ${\cal L}_{ik}v_k=0$ for $x_3\ge x_{3,R}$.
We store the  components $v_k({\bf x},\omega)$  in a $3\times 1$ vector ${\bf v}({\bf x},\omega)$ as follows
\begin{eqnarray}
{\bf v}({\bf x},\omega)=\begin{pmatrix} v_1\\v_2\\v_3\end{pmatrix}({\bf x},\omega).
\end{eqnarray}
In the lower half-space we express ${\bf v}({\bf x},\omega)$ as a superposition of 
mutually independent wave fields that obey the same source-free wave equation as ${\bf v}({\bf x},\omega)$ for $x_3\ge x_{3,R}$.
For this purpose we choose the focusing functions ${\bf F}({\bf x},{\bf x}_R,\omega)$ and ${\bf F}^*({\bf x},{\bf x}_R,\omega)$, of which the columns are also mutually independent. 
Hence, analogous to equation (\ref{eq12again}) we express ${\bf v}({\bf x},\omega)$ as
\begin{eqnarray}
{\bf v}({\bf x},\omega)&=&\int_{{{\partial\mathbb{D}}_R}} {\bf F}({\bf x},{\bf x}_R,\omega){\bf a}({\bf x}_R,\omega){\rm d}{\bf x}_R 
+\int_{{{\partial\mathbb{D}}_R}} {\bf F}^*({\bf x},{\bf x}_R,\omega){\bf b}({\bf x}_R,\omega){\rm d}{\bf x}_R, \nonumber\\
&&\hspace{8cm}\mbox{for}\quad x_3 \ge x_{3,R}. \label{eq318f}
\end{eqnarray}
Here ${\bf a}({\bf x}_R,\omega)$ and ${\bf b}({\bf x}_R,\omega)$ are as yet undetermined $3\times 1$ vectors.
In Appendix \ref{AppB1} we formulate  boundary conditions for the particle velocity and traction vectors at ${{\partial\mathbb{D}}_R}$, 
from which we solve ${\bf a}({\bf x}_R,\omega)$ and ${\bf b}({\bf x}_R,\omega)$. We thus obtain
\begin{eqnarray}
{\bf v}({\bf x},\omega)&=&\int_{{{\partial\mathbb{D}}_R}} {\bf F}({\bf x},{\bf x}_R,\omega){\bf v}^-({\bf x}_R,\omega){\rm d}{\bf x}_R 
+\int_{{{\partial\mathbb{D}}_R}} {\bf F}^*({\bf x},{\bf x}_R,\omega){\bf v}^+({\bf x}_R,\omega){\rm d}{\bf x}_R, \nonumber\\
&&\hspace{8cm}\mbox{for}\quad x_3 \ge x_{3,R}, \label{eq318}
\end{eqnarray}
where ${\bf v}^-({\bf x}_R,\omega)$ and ${\bf v}^+({\bf x}_R,\omega)$ represent the upgoing and downgoing parts, respectively, of ${\bf v}({\bf x}_R,\omega)$ for ${\bf x}_R$ at ${{\partial\mathbb{D}}_R}$. 
These upgoing and downgoing fields are velocity-normalized, meaning that ${\bf v}^-+{\bf v}^+={\bf v}$ at and above ${{\partial\mathbb{D}}_R}$.
Below  ${{\partial\mathbb{D}}_R}$ we only consider the total (undecomposed) wave field ${\bf v}$. 

The explanation of the right-hand side of equation (\ref{eq318})  in terms of Huygens' principle is similar to that of equation (\ref{eq12}).
The main extension is that the matrix-vector
products in equation (\ref{eq318}) accomplish a summation over the different components of the secondary sources at ${{{\partial\mathbb{D}}_R}}$
(corresponding to the foci of the different columns of the focusing function ${\bf F}$).

As for the acoustic representation of equation (\ref{eq12}), the  underlying assumption in the derivation of equation (\ref{eq318}) is that evanescent waves can be neglected at ${{\partial\mathbb{D}}_R}$.
Hence, it only holds for waves which have a horizontal slowness ${\bf s}$ which obeys 
\begin{eqnarray}
|{\bf s}|\le 1/c_P, \quad\mbox{at}\,\,{{\partial\mathbb{D}}_R}.\label{eqevanP}
\end{eqnarray}
Using similar arguments as given below equation (\ref{eqevan}), it follows that equation (\ref{eq318}) accounts for  evanescent waves inside the medium, as long as they are related
to propagating waves at the surface, as formulated by equation (\ref{eqevanP}).

\section{Elastodynamic Green's function representations}

\subsection{Representation for a modified elastodynamic Green's function}\label{sec5.1}

We introduce the elastodynamic Green's function $G_{k,n}({\bf x},{\bf x}_S,\omega)$ as a solution of equation (\ref{eqweedyn}) 
for a unit point source of volume-force density at ${\bf x}_S$ in the $x_n$-direction, hence
\begin{eqnarray}
{\cal L}_{ik}G_{k,n}=i\omega \delta_{in}\delta({\bf x}-{\bf x}_S).\label{eqweedynG}
\end{eqnarray}
We  demand that the time domain version of $G_{k,n}({\bf x},{\bf x}_S,\omega)$ is causal.
Note that $G_{k,n}$ obeys source-receiver reciprocity, i.e., $G_{k,n}({\bf x},{\bf x}_S,\omega)=G_{n,k}({\bf x}_S,{\bf x},\omega)$.
We introduce ${\bf G}({\bf x},{\bf x}_S,\omega)$ as a $3\times 3$ matrix, according to 
\begin{eqnarray}
{\bf G}({\bf x},{\bf x}_S,\omega)=
\begin{pmatrix}
G_{1,1} & G_{1,2} & G_{1,3} \\
G_{2,1} & G_{2,2} & G_{2,3} \\
G_{3,1} & G_{3,2} & G_{3,3} 
\end{pmatrix}({\bf x},{\bf x}_S,\omega).\label{eq319}
\end{eqnarray}
Each column is a particle velocity vector of which the components, 
for fixed ${\bf x}_S$ and variable ${\bf x}$, obey  wave equation (\ref{eqweedynG}). The different columns
correspond to different directions of the force source at ${\bf x}_S$.
This is different from the elastodynamic Green's function used by \cite{Wapenaar2014GJI}, in which the different elements represent decomposed compressional and shear waves.
In matrix form, source-receiver reciprocity implies ${\bf G}({\bf x},{\bf x}_S,\omega)=\{{\bf G}({\bf x}_S,{\bf x},\omega)\}^t$, where superscript $t$ denotes transposition.

We choose ${\bf x}_S=({\bf x}_{{\rm H},S},x_{3,S})$ again in the upper half-space, at a vanishing distance $\epsilon$ above ${{\partial\mathbb{D}}_R}$, hence, $x_{3,S}=x_{3,R}-\epsilon$.
In Appendix \ref{AppB2} we derive a modified version ${\bf \Gamma}({\bf x},{\bf x}_S,\omega)$ of ${\bf G}({\bf x},{\bf x}_S,\omega)$ (equation \ref{eq331}), of which the downgoing
part  ${\bf \Gamma}^+({\bf x},{\bf x}_S,\omega)$ for ${\bf x}$ at ${{\partial\mathbb{D}}_R}$ (i.e., just below the source level) is equal to a spatial delta function. Hence
\begin{eqnarray}
{\bf \Gamma}^+({\bf x},{\bf x}_S,\omega)|_{x_3=x_{3,R}}&=&{\bf I}\delta({\bf x}_{\rm H}-{\bf x}_{{\rm H},S}).\label{eq4Gag}
\end{eqnarray}
We define the reflection response ${\bf R}({\bf x}_R,{\bf x}_S,\omega)$ of the medium below ${{\partial\mathbb{D}}_R}$ as the upgoing part of ${\bf \Gamma}({\bf x}_R,{\bf x}_S,\omega)$,
 with ${\bf x}_R$ at ${{\partial\mathbb{D}}_R}$, hence
\begin{eqnarray}
{\bf R}({\bf x}_R,{\bf x}_S,\omega)&=&{\bf \Gamma}^-({\bf x}_R,{\bf x}_S,\omega).
\label{eq4Rag}
\end{eqnarray}
Substituting  ${\bf v}({\bf x},\omega)={\bf \Gamma}({\bf x},{\bf x}_S,\omega)$ and ${\bf v}^\pm({\bf x}_R,\omega)={\bf \Gamma}^\pm({\bf x}_R,{\bf x}_S,\omega)$ into equation (\ref{eq318}),
using equations (\ref{eq4Gag}) and (\ref{eq4Rag}), gives
 \begin{eqnarray}
{\bf \Gamma}({\bf x},{\bf x}_S,\omega)&=&\int_{{{\partial\mathbb{D}}_R}} {\bf F}({\bf x},{\bf x}_R,\omega){\bf R}({\bf x}_R,{\bf x}_S,\omega){\rm d}{\bf x}_R 
+{\bf F}^*({\bf x},{\bf x}_S,\omega),\quad\mbox{for}\quad x_3 \ge x_{3,R}.\label{eq339}
\end{eqnarray}
This is a representation for the modified version ${\bf \Gamma}({\bf x},{\bf x}_S,\omega)$ of the elastodynamic Green's function.
It has been derived without applying decomposition in the lower half-space.
It only excludes the contribution from waves that are evanescent at ${{\partial\mathbb{D}}_R}$.

\subsection{Representation for the  elastodynamic Green's function}\label{sec5.2}

In Appendix \ref{AppB3} we  show that equation (\ref{eq339}) can be reorganized into the following representation for the 
elastodynamic Green's function ${\bf G}({\bf x},{\bf x}_S,\omega)$
\begin{eqnarray}
{\bf G}({\bf x},{\bf x}_S,\omega)&=&\int_{{{\partial\mathbb{D}}_R}} {\bf f}({\bf x},{\bf x}_R,\omega)\{{\bf R}({\bf x}_S,{\bf x}_R,\omega)\}^t{\rm d}{\bf x}_R 
+{\bf f}^*({\bf x},{\bf x}_S,\omega), \quad\mbox{for}\quad x_3 \ge x_{3,R}.\label{eq349}
\end{eqnarray}
Here  ${\bf f}({\bf x},{\bf x}_R,\omega)$ is a modified version of the focusing function ${\bf F}({\bf x},{\bf x}_R,\omega)$ (equation \ref{eq341z}).
This representation gives the full elastodynamic particle velocity field at any virtual receiver position ${\bf x}$ inside the medium. 
It is similar to earlier derived elastodynamic representations for the Marchenko method \citep{Wapenaar2014GJI, Costa2014PRE}, 
but here it has been derived without applying decomposition at a truncation level inside the medium.
As a consequence, equation (\ref{eq349}) gives the full wave field at any virtual receiver position ${\bf x}$ inside the medium, 
including multiply reflected, converted, refracted and evanescent waves. This representation only excludes the contribution from waves that are evanescent at ${{\partial\mathbb{D}}_R}$, see 
the condition formulated by equation (\ref{eqevanP}). 

Applying elastodynamic representations like the one in equation (\ref{eq349}) to derive a Marchenko method is not trivial.
The functions ${\bf G}({\bf x},{\bf x}_S,\omega)$ and ${\bf f}^*({\bf x},{\bf x}_S,\omega)$, transformed back to the time domain, partly overlap and hence they cannot be 
completely separated by a time window (similar as discussed by \cite{Wapenaar2014GJI} and \cite{Reinicke2020GEO} for Green's functions and focusing functions consisting of
decomposed compressional and shear waves). A discussion of elastodynamic Marchenko methods is beyond the scope of this paper.

Similar as in the acoustic
situation, the representation of equation (\ref{eq349}) is not a sufficient starting point for imaging. We need at least  one other type of field at ${\bf x}$,
next to ${\bf G}({\bf x},{\bf x}_S,\omega)$, which represents the particle velocity at ${\bf x}$ in response to  force sources at ${\bf x}_S$.
To this end, we introduce a Green's function ${\bf G}_j^{\tau}({\bf x},{\bf x}_S,\omega)$ which, for $j=1, 2, 3$,  stands for the three traction vectors at ${\bf x}$.
From equation (\ref{eq434}) we derive that the traction vector ${\mbox{\boldmath $\tau$}}_j$ can be expressed in terms of 
the particle velocity as  ${\mbox{\boldmath $\tau$}}_j=-\frac{1}{i\omega}{\bf C}_{jl}\partial_l{\bf v}$, with $({\mbox{\boldmath $\tau$}}_j)_i=\tau_{ij}$ and $({\bf C}_{jl})_{ik}=c_{ijkl}$.
Similarly, we relate ${\bf G}_j^{\tau}$ to ${\bf G}$ via
\begin{eqnarray}
{\bf G}_j^{\tau}({\bf x},{\bf x}_S,\omega)=-\frac{1}{i\omega}{\bf C}_{jl}({\bf x})\partial_l{\bf G}({\bf x},{\bf x}_S,\omega).\label{eqtf}
\end{eqnarray}
Hence, when ${\bf G}({\bf x},{\bf x}_S,\omega)$ is available on a sufficiently dense grid, ${\bf G}_j^{\tau}({\bf x},{\bf x}_S,\omega)$ can be obtained via equation (\ref{eqtf}).
Alternatively, ${\bf G}_j^{\tau}({\bf x},{\bf x}_S,\omega)$ can be obtained from a modified version of the representation for ${\bf G}({\bf x},{\bf x}_S,\omega)$. 
Applying the operation $-\frac{1}{i\omega}{\bf C}_{jl}\partial_l$ to both sides of equation (\ref{eq349}) yields
\begin{eqnarray}
{\bf G}_j^{\tau}({\bf x},{\bf x}_S,\omega)&=&\int_{{{\partial\mathbb{D}}_R}} {\bf h}_j({\bf x},{\bf x}_R,\omega)\{{\bf R}({\bf x}_S,{\bf x}_R,\omega)\}^t{\rm d}{\bf x}_R 
-{\bf h}_j^*({\bf x},{\bf x}_S,\omega), \,\mbox{for}\quad x_3 \ge x_{3,R},\label{eq349g}
\end{eqnarray}
with
\begin{eqnarray}
{\bf h}_j({\bf x},{\bf x}_R,\omega)=-\frac{1}{i\omega}{\bf C}_{jl}({\bf x})\partial_l{\bf f}({\bf x},{\bf x}_R,\omega).
\end{eqnarray}
The Green's functions ${\bf G}({\bf x},{\bf x}_S,\omega)$ and ${\bf G}_j^{\tau}({\bf x},{\bf x}_S,\omega)$ together provide sufficient information for imaging.

\subsection{Representation for the homogeneous elastodynamic Green's function}\label{sec5.3}

The representations in sections \ref{sec5.1} and \ref{sec5.2} give the elastodynamic response to a source at ${\bf x}_S$,
observed by a virtual receiver at ${\bf x}$ inside the medium.
Similar as in section \ref{sec3.3}, here we modify the representation of equation (\ref{eq349}), to create the response at the surface to 
a virtual source inside the medium. 
After that, we show how to obtain the response to this virtual source at a virtual receiver inside the medium.

We start by renaming the coordinate vectors in equation (\ref{eq349}) as follows:
${\bf x}_S\to {\bf x}_R$,  ${\bf x}_R\to {\bf x}_S$, 
${\bf x}\to {{\bf x}_A}$. This yields, in combination with transposing all terms and 
applying source-receiver reciprocity on the left-hand side of equation (\ref{eq349}),
\begin{eqnarray}
{\bf G}({\bf x}_R,{{\bf x}_A},\omega)&=&\int_{{{\partial\mathbb{D}}_R}} {\bf R}({\bf x}_R,{\bf x}_S,\omega) {\bf f}^t({{\bf x}_A},{\bf x}_S,\omega){\rm d}{\bf x}_S 
+{\bf f}^\dagger({{\bf x}_A},{\bf x}_R,\omega), \,\mbox{for}\quad x_{3,A} \ge x_{3,R}.\label{eq350}
\end{eqnarray}
Here superscript $\dagger$ denotes transposition and complex conjugation.
The integral in  equation (\ref{eq350}) describes elastodynamic redatuming of the sources from all ${\bf x}_S$ at the surface to 
virtual-source position ${{\bf x}_A}$ in the subsurface.

Our next aim is to derive a representation
for the response observed by a virtual receiver at ${\bf x}$ in the subsurface, given ${\bf G}({\bf x}_R,{{\bf x}_A},\omega)$. 
Similar as in section \ref{sec3.3}, we define the homogeneous elastodynamic Green's function
 \begin{eqnarray}
{\bf G}_{\rm h}({\bf x},{{\bf x}_A},\omega)={\bf G}({\bf x},{{\bf x}_A},\omega)+{\bf G}^*({\bf x},{{\bf x}_A},\omega).
\end{eqnarray}
The components of the columns in ${\bf G}({\bf x},{{\bf x}_A},\omega)$ and ${\bf G}^*({\bf x},{{\bf x}_A},\omega)$ 
obey equation (\ref{eqweedynG}),
with source terms $i\omega\delta_{in}\delta({\bf x}-{{\bf x}_A})$ and $-i\omega\delta_{in}\delta({\bf x}-{{\bf x}_A})$, respectively, 
on the right-hand sides. 
Hence, the components of the columns of ${\bf G}_{\rm h}({\bf x},{{\bf x}_A},\omega)$ obey this equation without a source on
the right-hand side. Following a similar reasoning as in section \ref{sec3.3}, we substitute
\begin{eqnarray}
{\bf v}({\bf x},\omega)&=&{\bf G}_{\rm h}({\bf x},{{\bf x}_A},\omega),\\
{\bf v}^-({\bf x},\omega)&=&{\bf G}_{\rm h}^-({\bf x},{{\bf x}_A},\omega)=  {\bf G}({\bf x},{{\bf x}_A},\omega),\quad\mbox{for}\quad x_3=x_{3,R},\\
{\bf v}^+({\bf x},\omega)&=&{\bf G}_{\rm h}^+({\bf x},{{\bf x}_A},\omega)=  {\bf G}^*({\bf x},{{\bf x}_A},\omega),\quad\mbox{for}\quad x_3=x_{3,R},
\end{eqnarray}
into equation (\ref{eq318}). This gives
\begin{eqnarray}
{\bf G}_{\rm h}({\bf x},{{\bf x}_A},\omega)&=&2\Re\int_{{{\partial\mathbb{D}}_R}} {\bf F}({\bf x},{\bf x}_R,\omega){\bf G}({\bf x}_R,{{\bf x}_A},\omega){\rm d}{\bf x}_R,
\quad\mbox{for}\quad x_3 \ge x_{3,R}. \label{eq318re}
\end{eqnarray}
This equation describes elastodynamic redatuming of the receivers from all ${\bf x}_R$ at the surface to virtual-receiver position ${\bf x}$ in the subsurface.
It gives the homogeneous Green's function ${\bf G}_{\rm h}({\bf x},{{\bf x}_A},\omega)$, which is the response to the virtual source at ${{\bf x}_A}$, observed by
a virtual receiver at ${\bf x}$, plus its complex conjugate. After transforming this to the time domain,
 ${\bf G}({\bf x},{{\bf x}_A},t)$ can be extracted from ${\bf G}_{\rm h}({\bf x},{{\bf x}_A},t)$ by selecting  its causal part.

An elastodynamic homogeneous Green's function representation similar to equation (\ref{eq318re}) 
was also derived by \citet{Wapenaar2016RS} and illustrated with numerical examples by \citet{Reinicke2019WM},
 but here the fields are not decomposed into downgoing and upgoing compressional and shear waves inside the medium. 
Hence, it also holds for evanescent waves inside the medium, as long as condition (\ref{eqevanP}) is obeyed.
 Moreover, the derivation presented here is much simpler than in those references.

The source and receiver redatuming processes can be captured in one equation by substituting equation (\ref{eq350}) into  (\ref{eq318re}). This gives
\begin{eqnarray}
{\bf G}_{\rm h}({\bf x},{{\bf x}_A},\omega)&=&
2\Re\int_{{{\partial\mathbb{D}}_R}}\int_{{{\partial\mathbb{D}}_R}} {\bf F}({\bf x},{\bf x}_R,\omega){\bf R}({\bf x}_R,{\bf x}_S,\omega){\bf f}^t({{\bf x}_A},{\bf x}_S,\omega){\rm d}{\bf x}_S{\rm d}{\bf x}_R \nonumber\\
&+&2\Re\int_{{{\partial\mathbb{D}}_R}} {\bf F}({\bf x},{\bf x}_R,\omega){\bf f}^\dagger({{\bf x}_A},{\bf x}_R,\omega){\rm d}{\bf x}_R,
\quad\mbox{for}\quad \{x_3,x_{3,A}\}\ge x_{3,R}. \label{eq12abkel}
\end{eqnarray}
The double integral on the right-hand side resembles the process of classical elastodynamic source and receiver redatuming \citep{Wapenaar89Book, Hokstad2000GEO}, but with
the primary focusing functions in those references replaced by full-field focusing functions. It also resembles elastodynamic source-receiver interferometry \citep{Halliday2012GJI}, but with the
double integration along a closed boundary in that paper replaced by the double integration over the open boundary ${{\partial\mathbb{D}}_R}$.
Hence, via the theories of elastodynamic primary source-receiver redatuming \citep{Wapenaar89Book, Hokstad2000GEO}, closed-boundary source-receiver interferometry \citep{Halliday2012GJI}
and open-boundary homogeneous Green's function retrieval using wave field decomposition \citep{Wapenaar2016RS}, 
we have arrived at a representation for elastodynamic open-boundary homogeneous full-field Green's function retrieval  (equation \ref{eq12abkel}), 
which accounts for internal multiples, converted, refracted and evanescent waves in the lower half-space.

\section{Conclusions}

We have derived acoustic and elastodynamic Green's function representations in terms of the reflection response at the surface and focusing functions.
These representations have the same form as the representations that we derived earlier as the basis for Marchenko redatuming, imaging, monitoring and multiple elimination. 
However, unlike in our original derivations,
we did not assume that the wave field inside the medium can be decomposed into downgoing and upgoing waves and we did not ignore evanescent waves inside the medium.
We only neglected the contribution of waves that are evanescent at the acquisition boundary. 
We have demonstrated with numerical examples that the representations indeed account for evanescent waves inside the medium. 
The representations form a starting point for new research on  Marchenko methods which circumvent the  limitations caused by 
the  assumptions underlying the traditional representations. 
In these new developments,  care should be taken to account for the overlap in time of the Green's function and the time-reversed focusing function, 
particularly when dealing with refracted waves.

\section*{Acknowledgements}
We thank Marcin Dukalski, Mert Sinan Recep Kiraz and two anonymous reviewers for their comments, which helped us sharpen the message.
This work has received funding from the European Union's Horizon 2020 research and innovation programme: European Research Council (grant agreement 742703).



\bibliographystyle{gji}


\begin{thebibliography}{44}
\expandafter\ifx\csname natexlab\endcsname\relax\def\natexlab#1{#1}\fi

\bibitem[Behura et~al.(2014)Behura, Wapenaar, \& Snieder]{Behura2014GEO}
Behura, J., Wapenaar, K., \& Snieder, R., 2014.
\newblock Autofocus imaging: {I}mage reconstruction based on inverse scattering
  theory, {\it Geophysics\/}, {\bf 79}(3), A19--A26.

\bibitem[Berkhout(1982)]{Berkhout82Book}
Berkhout, A.~J., 1982.
\newblock {\it Seismic {M}igration. {I}maging of acoustic energy by wave field
  extrapolation. {A}. {T}heoretical aspects\/}, Elsevier.

\bibitem[Berryhill(1984)]{Berryhill84GEO}
Berryhill, J.~R., 1984.
\newblock Wave-equation datuming before stack, {\it Geophysics\/}, {\bf 49},
  2064--2066.

\bibitem[Brackenhoff et~al.(2019)Brackenhoff, Thorbecke, \&
  Wapenaar]{Brackenhoff2019SE}
Brackenhoff, J., Thorbecke, J., \& Wapenaar, K., 2019.
\newblock Monitoring of induced distributed double-couple sources using
  {M}archenko-based virtual receivers, {\it Solid Earth\/}, {\bf 10},
  1301--1319.

\bibitem[Broggini \& Snieder(2012)]{Broggini2012EJP}
Broggini, F. \& Snieder, R., 2012.
\newblock Connection of scattering principles: a visual and mathematical tour,
  {\it Eur. J. Phys.\/}, {\bf 33}, 593--613.

\bibitem[Broggini et~al.(2014)Broggini, Snieder, \& Wapenaar]{Broggini2014GEO}
Broggini, F., Snieder, R., \& Wapenaar, K., 2014.
\newblock Data-driven wavefield focusing and imaging with multidimensional
  deconvolution: {N}umerical examples for reflection data with internal
  multiples, {\it Geophysics\/}, {\bf 79}(3), WA107--WA115.

\bibitem[Burridge(1980)]{Burridge80WM}
Burridge, R., 1980.
\newblock The {G}elfand-{L}evitan, the {M}archenko, and the {G}opinath-{S}ondhi
  integral equations of inverse scattering theory, regarded in the context of
  inverse impulse-response problems, {\it Wave Motion\/}, {\bf 2}, 305--323.

\bibitem[Corones(1975)]{Corones75JMAA}
Corones, J.~P., 1975.
\newblock Bremmer series that correct parabolic approximations, {\it J.\ Math.\
  Anal.\ Appl.\/}, {\bf 50}, 361--372.

\bibitem[Curtis \& Halliday(2010)]{Curtis2010PRE}
Curtis, A. \& Halliday, D., 2010.
\newblock Source-receiver wavefield interferometry, {\it Phys. Rev. E\/},
  {\bf 81}, 046601.

\bibitem[da~Costa~Filho et~al.(2014)da~Costa~Filho, Ravasi, Curtis, \&
  Meles]{Costa2014PRE}
da~Costa~Filho, C.~A., Ravasi, M., Curtis, A., \& Meles, G.~A., 2014.
\newblock Elastodynamic {G}reen's function retrieval through single-sided
  {M}archenko inverse scattering, {\it Phys. Rev. E\/}, {\bf 90}, 063201.

\bibitem[Diekmann \& Vasconcelos(2021)]{Diekmann2021PRR}
Diekmann, L. \& Vasconcelos, I., 2021.
\newblock Focusing and {G}reen's function retrieval in three-dimensional
  inverse scattering revisited: {A} single-sided {M}archenko integral for the
  full wave field, {\it Phys. Rev. Research\/}, {\bf 3}, 013206.

\bibitem[Elison et~al.(2020)Elison, Dukalski, de~Vos, van Manen, \&
  Robertsson]{Elison2020GJI}
Elison, P., Dukalski, M.~S., de~Vos, K., van Manen, D.~J., \& Robertsson, J.
  O.~A., 2020.
\newblock Data-driven control over short-period internal multiples in media
  with a horizontally layered overburden, {\it Geophys. J. Int.\/}, {\bf 221}, 769--787.

\bibitem[Fishman \& McCoy(1984)]{Fishman84JMP}
Fishman, L. \& McCoy, J.~J., 1984.
\newblock Derivation and application of extended parabolic wave theories. {I}.
  {T}he factorized {H}elmholtz equation, {\it J.\ Math.\ Phys.\/}, {\bf 25}(2),
  285--296.

\bibitem[Halliday et~al.(2012)Halliday, Curtis, \& Wapenaar]{Halliday2012GJI}
Halliday, D., Curtis, A., \& Wapenaar, K., 2012.
\newblock Generalized {PP + PS = SS} from seismic interferometry, {\it
  Geophys. J. Int.\/}, {\bf 189}, 1015--1024.

\bibitem[Hokstad(2000)]{Hokstad2000GEO}
Hokstad, K., 2000.
\newblock Multicomponent {K}irchhoff migration, {\it Geophysics\/}, {\bf
  65}(3), 861--873.

\bibitem[Holicki et~al.(2019)Holicki, Drijkoningen, \& Wapenaar]{Holicki2019GP}
Holicki, M., Drijkoningen, G., \& Wapenaar, K., 2019.
\newblock Acoustic directional snapshot wavefield decomposition, {\it
  Geophys. Prosp.\/}, {\bf 67}, 32--51.

\bibitem[Jia et~al.(2018)Jia, Guitton, \& Snieder]{Jia2018GEO}
Jia, X., Guitton, A., \& Snieder, R., 2018.
\newblock A practical implementation of subsalt {M}archenko imaging with a
  {G}ulf of {M}exico data set, {\it Geophysics\/}, {\bf 83}(5), S409--S419.

\bibitem[Kennett et~al.(1978)Kennett, Kerry, \& Woodhouse]{Kennett78GJRAS}
Kennett, B. L.~N., Kerry, N.~J., \& Woodhouse, J.~H., 1978.
\newblock Symmetries in the reflection and transmission of elastic waves, {\it
  Geophys. J. R. astr. Soc.\/}, {\bf 52}, 215--230.

\bibitem[Kiraz et~al.(2021)Kiraz, Snieder, \& Wapenaar]{Kiraz2021JASA}
Kiraz, M. S.~R., Snieder, R., \& Wapenaar, K., 2021.
\newblock Focusing waves in an unknown medium without wavefield decomposition,
  {\it JASA Express Lett.\/}, {\bf 1}(5), 055602.
  
\bibitem[Liu et~al.(2011)Liu, Zhang, Morton, \& Leveille]{Liu2011GEO}
Liu, F., Zhang, G., Morton, S.~A., \& Leveille, J.~P., 2011.
\newblock An effective imaging condition for reverse-time migration using
  wavefield decomposition, {\it Geophysics\/}, {\bf 76}, S29--S39.

\bibitem[Lomas \& Curtis(2019)]{Lomas2019GEO}
Lomas, A. \& Curtis, A., 2019.
\newblock An introduction to {M}archenko methods for imaging, {\it
  Geophysics\/}, {\bf 84}(2), F35--F45.

\bibitem[Mildner et~al.(2019)Mildner, Broggini, de~Vos, \&
  Robertsson]{Mildner2019GEO}
Mildner, C., Broggini, F., de~Vos, K., \& Robertsson, J. O.~A., 2019.
\newblock Accurate source wavelet estimation using {M}archenko focusing
  functions, {\it Geophysics\/}, {\bf 84}(6), Q73--Q88.

\bibitem[Oristaglio(1989)]{Oristaglio89IP}
Oristaglio, M.~L., 1989.
\newblock An inverse scattering formula that uses all the data, {\it Inverse Probl.\/}, {\bf 5}, 1097--1105.

\bibitem[Pereira et~al.(2019)Pereira, Ramzy, Griscenco, Huard, Huang, Cypriano,
  \& Khalil]{Pereira2019SEG}
Pereira, R., Ramzy, M., Griscenco, P., Huard, B., Huang, H., Cypriano, L., \&
  Khalil, A., 2019.
\newblock Internal multiple attenuation for {OBN} data with overburden/target
  separation, in {\em Proceedings of the 89th Annual Meeting of the Society of Exploration Geophysicists\/}, pp. 4520--4524.

\bibitem[Porter(1970)]{Porter70JOSA}
Porter, R.~P., 1970.
\newblock Diffraction-limited, scalar image formation with holograms of
  arbitrary shape, {\it J. opt. Soc. Am.\/}, {\bf 60},
  1051--1059.

\bibitem[Ravasi \& Vasconcelos(2021)]{Ravasi2021GEO}
Ravasi, M. \& Vasconcelos, I., 2021.
\newblock An open-source framework for the implementation of large-scale
  integral operators with flexible, modern {HPC} solutions - enabling 3{D
  M}archenko imaging by least squares inversion, {\it Geophysics\/}, {\bf
  86}(early view), doi.org/10.1190/geo2020--0796.1.

\bibitem[Ravasi et~al.(2016)Ravasi, Vasconcelos, Kritski, Curtis,
  da~Costa~Filho, \& Meles]{Ravasi2016GJI}
Ravasi, M., Vasconcelos, I., Kritski, A., Curtis, A., da~Costa~Filho, C.~A., \&
  Meles, G.~A., 2016.
\newblock Target-oriented {M}archenko imaging of a {N}orth {S}ea field, {\it
  Geophys. J. Int.\/}, {\bf 205}, 99--104.

\bibitem[Reinicke \& Wapenaar(2019)]{Reinicke2019WM}
Reinicke, C. \& Wapenaar, K., 2019.
\newblock Elastodynamic single-sided homogeneous {G}reen's function
  representation: {T}heory and numerical examples, {\it Wave Motion\/}, {\bf
  89}, 245--264.

\bibitem[Reinicke et~al.(2020)Reinicke, Dukalski, \& Wapenaar]{Reinicke2020GEO}
Reinicke, C., Dukalski, M., \& Wapenaar, K., 2020.
\newblock Comparison of monotonicity challenges encountered by the inverse
  scattering series and the {M}archenko demultiple method for elastic waves,
  {\it Geophysics\/}, {\bf 85}(5), Q11--Q26.

\bibitem[Schoenberg \& Sen(1983)]{Schoenberg83JASA}
Schoenberg, M. \& Sen, P.~N., 1983.
\newblock Properties of a periodically stratified acoustic half-space and its
  relation to a {B}iot fluid, {\it J. acoust. Soc. Am.\/}, {\bf 73}, 61--67.

\bibitem[Singh \& Snieder(2017)]{Singh2017GEO2}
Singh, S. \& Snieder, R., 2017.
\newblock Source-receiver {M}archenko redatuming: {O}btaining virtual receivers
  and virtual sources in the subsurface, {\it Geophysics\/}, {\bf 82}(3),
  Q13--Q21.

\bibitem[Slob et~al.(2014)Slob, Wapenaar, Broggini, \& Snieder]{Slob2014GEO}
Slob, E., Wapenaar, K., Broggini, F., \& Snieder, R., 2014.
\newblock Seismic reflector imaging using internal multiples with
  {M}archenko-type equations, {\it Geophysics\/}, {\bf 79}(2), S63--S76.

\bibitem[Staring \& Wapenaar(2020)]{Staring2020GP}
Staring, M. \& Wapenaar, K., 2020.
\newblock Three-dimensional {M}archenko internal multiple attenuation on narrow
  azimuth streamer data of the {S}antos {B}asin, {B}razil, {\it Geophys.
  Prosp.\/}, {\bf 68}, 1864--1877.

\bibitem[Staring et~al.(2018)Staring, Pereira, Douma, van~der Neut, \&
  Wapenaar]{Staring2018GEO}
Staring, M., Pereira, R., Douma, H., van~der Neut, J., \& Wapenaar, K., 2018.
\newblock Source-receiver {M}archenko redatuming on field data using an
  adaptive double-focusing method, {\it Geophysics\/}, {\bf 83}(6), S579--S590.

\bibitem[Stoffa(1989)]{Stoffa89Book}
Stoffa, P.~L., 1989.
\newblock {\it Tau-p - {A} plane wave approach to the analysis of seismic
  data\/}, Kluwer Academic Publishers, Dordrecht.

\bibitem[Ursin(1983)]{Ursin83GEO}
Ursin, B., 1983.
\newblock Review of elastic and electromagnetic wave propagation in
  horizontally layered media, {\it Geophysics\/}, {\bf 48}, 1063--1081.

\bibitem[van~der Neut et~al.(2017)Van~der Neut, Johnson, van Wijk, Singh, Slob,
  \& Wapenaar]{Neut2017JASA}
Van~der Neut, J., Johnson, J.~L., van Wijk, K., Singh, S., Slob, E., \&
  Wapenaar, K., 2017.
\newblock A {M}archenko equation for acoustic inverse source problems, {\it
  J. acoust. Soc. Am.\/}, {\bf 141}(6), 4332--4346.

\bibitem[Wapenaar \& Berkhout(1989)]{Wapenaar89Book}
Wapenaar, C. P.~A. \& Berkhout, A.~J., 1989.
\newblock {\it Elastic wave field extrapolation\/}, Elsevier, Amsterdam.

\bibitem[Wapenaar(2020)]{Wapenaar2020GJI}
Wapenaar, K., 2020.
\newblock The {M}archenko method for evanescent waves, {\it Geophys. J. Int.\/}, {\bf 223}, 1412--1417.

\bibitem[Wapenaar \& Slob(2014)]{Wapenaar2014GJI}
Wapenaar, K. \& Slob, E., 2014.
\newblock On the {M}archenko equation for multicomponent single-sided
  reflection data, {\it Geophys. J. Int.\/}, {\bf 199},
  1367--1371.

\bibitem[Wapenaar et~al.(2013)Wapenaar, Broggini, Slob, \&
  Snieder]{Wapenaar2013PRL}
Wapenaar, K., Broggini, F., Slob, E., \& Snieder, R., 2013.
\newblock Three-dimensional single-sided {M}archenko inverse scattering,
  data-driven focusing, {G}reen's function retrieval, and their mutual
  relations, {\it Phys. Rev. Lett.\/}, {\bf 110}, 084301.

\bibitem[Wapenaar et~al.(2014)Wapenaar, Thorbecke, van~der Neut, Broggini,
  Slob, \& Snieder]{Wapenaar2014GEO}
Wapenaar, K., Thorbecke, J., van~der Neut, J., Broggini, F., Slob, E., \&
  Snieder, R., 2014.
\newblock Marchenko imaging, {\it Geophysics\/}, {\bf 79}(3), WA39--WA57.

\bibitem[Wapenaar et~al.(2016{\natexlab{a}})Wapenaar, Thorbecke, \& van~der
  Neut]{Wapenaar2016GJI}
Wapenaar, K., Thorbecke, J., \& van~der Neut, J., 2016{\natexlab{a}}.
\newblock A single-sided homogeneous {G}reen's function representation for
  holographic imaging, inverse scattering, time-reversal acoustics and
  interferometric {G}reen's function retrieval, {\it Geophys. J. Int.\/}, {\bf 205}, 531--535.

\bibitem[Wapenaar et~al.(2016{\natexlab{b}})Wapenaar, van~der Neut, \&
  Slob]{Wapenaar2016RS}
Wapenaar, K., van~der Neut, J., \& Slob, E., 2016{\natexlab{b}}.
\newblock Unified double- and single-sided homogeneous {G}reen's function
  representations, {\it Proc. R. Soc. A\/}, {\bf 472},
  20160162.

\bibitem[Yoon \& Marfurt(2006)]{Yoon2006EG}
Yoon, K. \& Marfurt, K.~J., 2006.
\newblock Reverse-time migration using the {P}oynting vector, {\it Exploration
  Geophysics\/}, {\bf 37}, 102--107.

\bibitem[Zhang \& Slob(2020)]{Zhang2020GJI}
Zhang, L. \& Slob, E., 2020.
\newblock A fast algorithm for multiple elimination and transmission
  compensation in primary reflections, {\it Geophys. J. Int.\/}, {\bf 221}, 371--377.

\end{thebibliography}

\appendix

\section{Derivation of the acoustic wave field representation}

\subsection{Derivation of the representation of equation (\ref{eq12})}\label{AppA1}

We derive expressions for the coefficients $a({\bf x}_R,\omega)$ and $b({\bf x}_R,\omega)$ in the acoustic wave field representation of equation (\ref{eq12again}).
We do this by formulating two boundary conditions at ${{\partial\mathbb{D}}_R}$.
First, we consider the acoustic pressure $p({\bf x},\omega)$ at ${{\partial\mathbb{D}}_R}$. To this end, we  evaluate equation (\ref{eq12again}) for ${\bf x}$ at ${{\partial\mathbb{D}}_R}$. 
Using the focusing condition formulated in equation (\ref{eq4}) we thus obtain
\begin{eqnarray}
p({\bf x},\omega)|_{x_3=x_{3,R}}&=&\int_{{{\partial\mathbb{D}}_R}} \delta({\bf x}_{\rm H}-{\bf x}_{{\rm H},R})a({\bf x}_R,\omega){\rm d}{\bf x}_R 
+\int_{{{\partial\mathbb{D}}_R}} \delta({\bf x}_{\rm H}-{\bf x}_{{\rm H},R})b({\bf x}_R,\omega){\rm d}{\bf x}_R, \nonumber\\
&=&\{a({\bf x},\omega)+b({\bf x},\omega)\}_{x_3=x_{3,R}},\label{eq12againd}
\end{eqnarray}
where we used ${\bf x}_R=({\bf x}_{{\rm H},R},x_{3,R})$. This is our first equation for the coefficients $a({\bf x}_R,\omega)$ and $b({\bf x}_R,\omega)$.

Next, we consider the vertical component of the particle velocity $v_3({\bf x},\omega)$ at ${{\partial\mathbb{D}}_R}$. 
From the Fourier transform of equation (\ref{eqbeq1}), using $\rho_{jk}=\delta_{jk}\rho_0$ at ${{\partial\mathbb{D}}_R}$, we obtain
$v_3({\bf x},\omega)=\frac{1}{i\omega\rho_0}\partial_3p({\bf x},\omega)$ for ${\bf x}$ at ${{\partial\mathbb{D}}_R}$. Substituting equation (\ref{eq12again}) gives
\begin{eqnarray}
v_3({\bf x},\omega)&=&\frac{1}{i\omega\rho_0}\int_{{{\partial\mathbb{D}}_R}} \partial_3 F({\bf x},{\bf x}_R,\omega)a({\bf x}_R,\omega){\rm d}{\bf x}_R
 \nonumber\\&+&
 \frac{1}{i\omega\rho_0}\int_{{{\partial\mathbb{D}}_R}} \partial_3 F^*({\bf x},{\bf x}_R,\omega)b({\bf x}_R,\omega){\rm d}{\bf x}_R, \label{eq12derivative}
\end{eqnarray}
for $x_3=x_{3,R}$.
Applying the spatial Fourier transformation of equation (\ref{eq50a}) to both sides of equation (\ref{eq12derivative}) gives
\begin{eqnarray}
\tilde v_3({\bf s},x_3,\omega)&=&
\frac{1}{i\omega\rho_0}\int_{{{\partial\mathbb{D}}_R}} \partial_3\tilde F({\bf s},x_3,{\bf x}_R,\omega)a({\bf x}_R,\omega){\rm d}{\bf x}_R 
\nonumber\\&+&
\frac{1}{i\omega\rho_0}\int_{{{\partial\mathbb{D}}_R}} \partial_3\tilde F^*(-{\bf s},x_3,{\bf x}_R,\omega)b({\bf x}_R,\omega){\rm d}{\bf x}_R, \label{eq12againn}
\end{eqnarray}
for $x_3=x_{3,R}$. At this depth level the focusing function is an upgoing field (see Figure \ref{Figure1}), hence it obeys the following one-way wave equation  
\begin{eqnarray}
&&\hspace{-0.5cm}\partial_3\tilde F({\bf s},x_3,{\bf x}_R,\omega)|_{x_3=x_{3,R}}= -i\omega s_3\tilde F({\bf s},x_{3,R},{\bf x}_R,\omega),\label{eq54pr}
\end{eqnarray}
 with the vertical slowness $s_3$ defined as 
\begin{eqnarray}
s_3=\begin{cases}
\sqrt{1/c_0^2-{\bf s}\cdot{\bf s}}, &\mbox{for } {\bf s}\cdot{\bf s}\le 1/c_0^2\\
i\sqrt{{\bf s}\cdot{\bf s}-1/c_0^2}, &\mbox{for } {\bf s}\cdot{\bf s}> 1/c_0^2.
\end{cases}\label{eq53prsq}
\end{eqnarray}
The two expressions  in equation (\ref{eq53prsq}) represent the situation for propagating and evanescent waves, respectively.
Applying the spatial Fourier transformation of equation (\ref{eq50a}) to   equation (\ref{eq4})
we further have
\begin{eqnarray}
\tilde F({\bf s},x_{3,R},{\bf x}_R,\omega)=\exp\{-i\omega{\bf s}\cdot{\bf x}_{{\rm H},R}\}. \label{eq55pr}
\end{eqnarray}
Substitution of equations (\ref{eq54pr}) and (\ref{eq55pr}) into equation (\ref{eq12againn}) for $x_3=x_{3,R}$  gives
\begin{eqnarray}
&&\tilde v_3({\bf s},x_{3,R},\omega)\nonumber\\
&&=-\frac{s_3}{\rho_0}\int_{{{\partial\mathbb{D}}_R}} \exp\{-i\omega{\bf s}\cdot{\bf x}_{{\rm H},R}\}a({\bf x}_R,\omega){\rm d}{\bf x}_R 
+\frac{s_3^*}{\rho_0}\int_{{{\partial\mathbb{D}}_R}} \exp\{-i\omega{\bf s}\cdot{\bf x}_{{\rm H},R}\}b({\bf x}_R,\omega){\rm d}{\bf x}_R \nonumber\\
&&=-\frac{s_3}{\rho_0}\tilde a({\bf s},x_{3,R},\omega)+\frac{s_3^*}{\rho_0}\tilde b({\bf s},x_{3,R},\omega).\label{eq12againnn}
\end{eqnarray}
Combining the spatial Fourier transform of equation (\ref{eq12againd}) with equation (\ref{eq12againnn}) gives
\begin{eqnarray}
\begin{pmatrix} \tilde p\\ \tilde v_3\end{pmatrix}_{x_3=x_{3,R}}=
\begin{pmatrix} 1 & 1 \\ s_3^*/\rho_0 & -s_3/\rho_0\end{pmatrix}
\begin{pmatrix} \tilde b\\ \tilde a\end{pmatrix}_{x_3=x_{3,R}}.\label{eq57pqrs}
\end{eqnarray}
For ${\bf s}\cdot{\bf s} \le 1/c_0^2$ at ${{\partial\mathbb{D}}_R}$ we have $s_3^*=s_3$, see equation (\ref{eq53prsq}). Hence, for propagating waves, equation (\ref{eq57pqrs}) is recognised as
 the well-known system that
composes the total wave fields on the left-hand side from downgoing and upgoing fields on the right-hand side 
\citep{Corones75JMAA, Ursin83GEO, Fishman84JMP}. Hence
\begin{eqnarray}
\tilde b({\bf s},x_{3,R},\omega)&=&\tilde p^+({\bf s},x_{3,R},\omega),\label{eq67prs}\\
\tilde a({\bf s},x_{3,R},\omega)&=&\tilde p^-({\bf s},x_{3,R},\omega),\label{eq67pr}
\end{eqnarray}
for ${\bf s}\cdot{\bf s} \le 1/c_0^2$ at ${{\partial\mathbb{D}}_R}$, where $\tilde p^+({\bf s},x_{3,R},\omega)$ and $\tilde p^-({\bf s},x_{3,R},\omega)$ are downgoing and upgoing fields, 
respectively, at ${{\partial\mathbb{D}}_R}$.
Transforming these expressions back to the space domain, using
\begin{eqnarray}
&&\hspace{-0.7cm}  p^\pm({\bf x}_R,\omega)=\frac{\omega^2}{4\pi^2}\int_{\mathbb{R}^2}\exp\{i\omega{\bf s}\cdot{\bf x}_{{\rm H},R}\}\tilde p^\pm({\bf s},x_{3,R},\omega){\rm d}{\bf s}\label{eq50aprs}
\end{eqnarray} 
and  similar expressions for $a({\bf x}_R,\omega)$ and $b({\bf x}_R,\omega)$, 
 gives
\begin{eqnarray}
b({\bf x}_R,\omega) &\approx& p^+({\bf x}_R,\omega),\label{eqa13}\\
a({\bf x}_R,\omega) &\approx& p^-({\bf x}_R,\omega).\label{eqa14}
\end{eqnarray}
The approximation signs signify that evanescent waves are neglected at ${{\partial\mathbb{D}}_R}$
 (since equations (\ref{eq67prs}) and (\ref{eq67pr}) hold for propagating waves only,
 whereas the inverse Fourier transformation involves an integration along all horizontal slownesses).
Substitution of equations (\ref{eqa13}) and (\ref{eqa14}) into equation (\ref{eq12again}) gives equation (\ref{eq12}).

\subsection{Analysis of the integral in equation (\ref{eq56h})}\label{AppA2}

We analyze the integral in equation (\ref{eq56h}). 
We show that we can transfer the operator $\partial_{3,R}$  from $f$ to $G^{\rm s}$, and that this is accompanied with a sign change.
For a function  of two space variables, $u({\bf x},{\bf x}_R,\omega)$, we define the spatial Fourier transform along the second space variable as
\begin{eqnarray}
&&\tilde u({\bf x},{\bf s},x_{3,R},\omega)=
\int_{\mathbb{R}^2}u({\bf x},{\bf x}_{{\rm H},R},x_{3,R},\omega)\exp\{i\omega{\bf s}\cdot{\bf x}_{{\rm H},R}\}{\rm d}{\bf x}_{{\rm H},R}.\label{eq329}
\end{eqnarray}
Note the opposite sign in the exponential, compared with that in equation (\ref{eq50a}).
Using this Fourier transform and Parseval's theorem, we obtain for the  integral in equation (\ref{eq56h})
\begin{eqnarray}
&&\hspace{-0.7cm}\int_{{{\partial\mathbb{D}}_R}} \{\partial_{3,R} f({\bf x},{\bf x}_R,\omega)\}G^{\rm s}({\bf x}_S,{\bf x}_R,\omega){\rm d}{\bf x}_R=\nonumber\\
&&\hspace{-0.7cm}\frac{\omega^2}{4\pi^2}\int_{\mathbb{R}^2} \{\partial_{3,R} \tilde f({\bf x},-{\bf s},x_{3,R},\omega)\}\tilde G^{\rm s}({\bf x}_S,{\bf s},x_{3,R},\omega){\rm d}{\bf s}.\label{eq15P}
\end{eqnarray}
Note that $\tilde f({\bf x},-{\bf s},x_{3,R},\omega)$ is differentiated with respect to the focal point depth $x_{3,R}$, hence, the one-way wave equation gets a sign opposite to that in
equation (\ref{eq54pr}), i.e. $\partial_{3,R} \tilde f({\bf x},-{\bf s},x_{3,R},\omega)=i\omega s_3\tilde f({\bf x},-{\bf s},x_{3,R},\omega)$, with $s_3$ defined in equation (\ref{eq53prsq}).
We transfer $i\omega s_3$ to the Green's function and use $i\omega s_3\tilde G^{\rm s}({\bf x}_S,{\bf s},x_{3,R},\omega)=-\partial_{3,R}\tilde G^{\rm s}({\bf x}_S,{\bf s},x_{3,R},\omega)$
(which is a differentiation with respect to the source depth $x_{3,R}$). Making these substitutions in the right-hand side of equation (\ref{eq15P}) and 
applying Parseval's theorem again gives
\begin{eqnarray}
&&\hspace{-0.7cm}-\frac{\omega^2}{4\pi^2}\int_{\mathbb{R}^2} \tilde f({\bf x},-{\bf s},x_{3,R},\omega)\partial_{3,R}\tilde G^{\rm s}({\bf x}_S,{\bf s},x_{3,R},\omega){\rm d}{\bf s}\nonumber\\
&&\hspace{-0.7cm}=-\int_{{{\partial\mathbb{D}}_R}}  f({\bf x},{\bf x}_R,\omega)\partial_{3,R} G^{\rm s}({\bf x}_S,{\bf x}_R,\omega){\rm d}{\bf x}_R.
\end{eqnarray}
Hence, we have transferred the operator $\partial_{3,R}$ under the integral in equation (\ref{eq56h}) from $f$ to $G^{\rm s}$, which involves a sign change.

\section{Derivation of the elastodynamic wave field representation}

\subsection{Derivation of the  representation of equation (\ref{eq318})}\label{AppB1}

We derive expressions for the coefficients ${\bf a}({\bf x}_R,\omega)$ and ${\bf b}({\bf x}_R,\omega)$ in the elastodynamic wave field representation of equation (\ref{eq318f}).
We do this by formulating two boundary conditions at ${{\partial\mathbb{D}}_R}$.
First, we consider the particle velocity vector ${\bf v}({\bf x},\omega)$ at ${{\partial\mathbb{D}}_R}$. To this end, we  evaluate equation (\ref{eq318f}) for ${\bf x}$ at ${{\partial\mathbb{D}}_R}$. 
Using the focusing condition formulated in equation (\ref{eq302}) we thus obtain
\begin{eqnarray}
{\bf v}({\bf x},\omega)|_{x_3=x_{3,R}}&=&\int_{{{\partial\mathbb{D}}_R}} \delta({\bf x}_{\rm H}-{\bf x}_{{\rm H},R}){\bf a}({\bf x}_R,\omega){\rm d}{\bf x}_R 
+\int_{{{\partial\mathbb{D}}_R}} \delta({\bf x}_{\rm H}-{\bf x}_{{\rm H},R}){\bf b}({\bf x}_R,\omega){\rm d}{\bf x}_R, 
\nonumber\\&=&
\{{\bf a}({\bf x},\omega)+{\bf b}({\bf x},\omega)\}_{x_3=x_{3,R}}.\label{eq304}
\end{eqnarray}
For the second boundary condition we analyze the traction vector ${\mbox{\boldmath $\tau$}}_3({\bf x},\omega)$ at  ${{\partial\mathbb{D}}_R}$. 
First we establish a relation between ${\mbox{\boldmath $\tau$}}_3({\bf x},\omega)$ and ${\bf v}({\bf x},\omega)$ 
in the homogeneous isotropic upper half-space, including ${{\partial\mathbb{D}}_R}$.
Using equation (\ref{eq50a}), we transform ${\bf v}({\bf x},\omega)$ and ${\mbox{\boldmath $\tau$}}_3({\bf x},\omega)$ for $x_3\le x_{3,R}$ 
to $\tilde{\bf v}({\bf s},x_3,\omega)$ and $\tilde{\mbox{\boldmath $\tau$}}_3({\bf s},x_3,\omega)$, respectively.
These fields can be related to vectors $\tilde{\bf p}^+$ and $\tilde{\bf p}^-$ containing downgoing and upgoing compressional and shear wave fields, according to
\begin{eqnarray}
\begin{pmatrix} \tilde{\bf v} \\-\tilde{\mbox{\boldmath $\tau$}}_3 \end{pmatrix} =
\begin{pmatrix} \tilde{\bf L}_1^+ & \tilde{\bf L}_1^- \\ \tilde{\bf L}_2^+ & \tilde{\bf L}_2^- \end{pmatrix}
\begin{pmatrix} \tilde{\bf p}^+\\\tilde{\bf p}^- \end{pmatrix},\quad\mbox{for}\quad x_3\le x_{3,R},\label{eq305k}
\end{eqnarray}
\citep{Kennett78GJRAS, Ursin83GEO, Wapenaar89Book}. Next, we define the downgoing and upgoing parts of $\tilde{\bf v}$ as 
$\tilde{\bf v}^\pm= \tilde{\bf L}_1^\pm\tilde{\bf p}^\pm$ 
and rewrite equation (\ref{eq305k}) as
\begin{eqnarray}
\begin{pmatrix} \tilde{\bf v}\\-\tilde{\mbox{\boldmath $\tau$}}_3 \end{pmatrix} =\begin{pmatrix} {\bf I} & {\bf I}\\\tilde{\bf D}^+ & \tilde{\bf D}^-  \end{pmatrix}
\begin{pmatrix} \tilde{\bf v}^+\\\tilde{\bf v}^- \end{pmatrix},\quad\mbox{for}\quad x_3\le x_{3,R},\label{eq305}
\end{eqnarray}
with  $\tilde{\bf D}^\pm= \tilde{\bf L}_2^\pm(\tilde{\bf L}_1^\pm)^{-1}$. 
The matrices $\tilde{\bf L}_1^\pm$ and $\tilde{\bf L}_2^\pm$ in equation (\ref{eq305k})
are not uniquely defined.
They depend on the chosen normalization of the fields contained in $\tilde{\bf p}^+$ and $\tilde{\bf p}^-$. However, independent of the  normalization,
the matrix $\tilde{\bf D}^\pm$ in equation (\ref{eq305}) is uniquely defined. It is given by
\begin{eqnarray}
&&\hspace{-.7cm}\tilde{\bf D}^\pm({\bf s})=
\\&&\hspace{-.7cm}
\frac{\rho_0 c_S^2}{s_3^P s_3^S +{\bf s}\cdot{\bf s}}\begin{pmatrix} \pm((c_S^{-2}-s_2^2)s_3^P + s_2^2s_3^S) & \pm s_1s_2(s_3^P -s_3^S) & -s_1(c_S^{-2}-2(s_3^P s_3^S+{\bf s}\cdot{\bf s}))\\
 \pm s_1s_2(s_3^P -s_3^S) & \pm((c_S^{-2}-s_1^2)s_3^P + s_1^2s_3^S)&  -s_2(c_S^{-2}-2(s_3^P s_3^S+{\bf s}\cdot{\bf s}))\\
 s_1(c_S^{-2}-2(s_3^P s_3^S+{\bf s}\cdot{\bf s})) &s_2(c_S^{-2}-2(s_3^P s_3^S+{\bf s}\cdot{\bf s})) &\pm s_3^S c_S^{-2}
\end{pmatrix}
\nonumber
\end{eqnarray}
with the vertical slownesses $s_3^P$ and $s_3^S$ for $P$- and $S$-waves, respectively, defined as
\begin{eqnarray}
s_3^{P,S}=\begin{cases}
\sqrt{1/c_{P,S}^2-{\bf s}\cdot{\bf s}}, &\mbox{for } {\bf s}\cdot{\bf s}\le 1/c_{P,S}^2\\
i\sqrt{{\bf s}\cdot{\bf s}-1/c_{P,S}^2}, &\mbox{for } {\bf s}\cdot{\bf s}> 1/c_{P,S}^2,
\end{cases}\label{eq324}
\end{eqnarray}
where $c_P$ and $c_S$ are the $P$- and $S$-wave velocities, respectively, of the upper half-space $x_3\le x_{3,R}$.
Applying the transform of equation (\ref{eq50a}) to equation (\ref{eq318f}) we obtain for ${\bf x}$ at  ${{\partial\mathbb{D}}_R}$
\begin{eqnarray}
\tilde{\bf v}({\bf s},x_{3,R},\omega)&=&\int_{{{\partial\mathbb{D}}_R}} \tilde{\bf F}({\bf s},x_{3,R},{\bf x}_R,\omega){\bf a}({\bf x}_R,\omega){\rm d}{\bf x}_R 
+\int_{{{\partial\mathbb{D}}_R}} \tilde{\bf F}^*(-{\bf s},x_{3,R},{\bf x}_R,\omega){\bf b}({\bf x}_R,\omega){\rm d}{\bf x}_R. \nonumber\\
&&\label{eq308}
\end{eqnarray}
Since $\tilde{\bf F}({\bf s},x_{3,R},{\bf x}_R,\omega)$ is upgoing,
the first term on the right-hand side is the upgoing velocity field $\tilde{\bf v}^-({\bf s},x_{3,R},\omega)$ and the second term is, for propagating waves
(i.e., for  ${\bf s}\cdot{\bf s}\le 1/c_P^2$), the downgoing velocity field  $\tilde{\bf v}^+({\bf s},x_{3,R},\omega)$.
Hence, using equation (\ref{eq305}) we obtain for the transformed traction vector
\begin{eqnarray}
&&-\tilde{\mbox{\boldmath $\tau$}}_3({\bf s},x_{3,R},\omega)=\nonumber\\
&&\hspace{0.5cm}\tilde{\bf D}^-({\bf s})\int_{{{\partial\mathbb{D}}_R}}  \tilde{\bf F}({\bf s},x_{3,R},{\bf x}_R,\omega){\bf a}({\bf x}_R,\omega){\rm d}{\bf x}_R 
\nonumber\\&&\hspace{0.5cm}
+\tilde{\bf D}^+({\bf s})\int_{{{\partial\mathbb{D}}_R}}  \tilde{\bf F}^*(-{\bf s},x_{3,R},{\bf x}_R,\omega){\bf b}({\bf x}_R,\omega){\rm d}{\bf x}_R, 
\label{eq309}
\end{eqnarray}
for ${\bf s}\cdot{\bf s}\le 1/c_P^2$ at ${{\partial\mathbb{D}}_R}$.
Applying the transform of equation (\ref{eq50a}) to the focusing condition of equation (\ref{eq302}) gives
\begin{eqnarray}
\tilde {\bf F}({\bf s},x_{3,R},{\bf x}_R,\omega)={\bf I}\exp\{-i\omega {\bf s}\cdot{\bf x}_{{\rm H},R}\}. \label{eq310}
\end{eqnarray}
Substituting this into equation (\ref{eq309}) we obtain
\begin{eqnarray}
&&-\tilde{\mbox{\boldmath $\tau$}}_3({\bf s},x_{3,R},\omega)=
\tilde{\bf D}^-({\bf s}) \tilde{\bf a}({\bf s},x_{3,R},\omega)+ \tilde{\bf D}^+({\bf s}) \tilde{\bf b}({\bf s},x_{3,R},\omega),\label{eq311}
\end{eqnarray}
for ${\bf s}\cdot{\bf s}\le 1/c_P^2$ at ${{\partial\mathbb{D}}_R}$. Combining this equation with the Fourier transform of equation (\ref{eq304}) yields
\begin{eqnarray}
\begin{pmatrix} \tilde{\bf v}\\-\tilde{\mbox{\boldmath $\tau$}}_3 \end{pmatrix}_{x_3=x_{3,R}} &=&\begin{pmatrix}  {\bf I} & {\bf I}\\\tilde{\bf D}^+ & \tilde{\bf D}^- \end{pmatrix}
\begin{pmatrix} \tilde{\bf b}\\\tilde{\bf a} \end{pmatrix}_{x_3=x_{3,R}},\label{eq312}
\end{eqnarray}
for  ${\bf s}\cdot{\bf s}\le 1/c_P^2$ at ${{\partial\mathbb{D}}_R}$. Comparing this with equation (\ref{eq305}) we conclude
\begin{eqnarray}
\tilde{\bf b}({\bf s},x_{3,R},\omega)&=&\tilde {\bf v}^+({\bf s},x_{3,R},\omega),\label{eq313}\\
\tilde{\bf a}({\bf s},x_{3,R},\omega)&=&\tilde {\bf v}^-({\bf s},x_{3,R},\omega),\label{eq314}
\end{eqnarray}
for  ${\bf s}\cdot{\bf s}\le 1/c_P^2$ at ${{\partial\mathbb{D}}_R}$. 
 Transforming these expressions back to the space domain gives
\begin{eqnarray}
{\bf b}({\bf x}_R,\omega) &\approx& {\bf v}^+({\bf x}_R,\omega),\label{eq315}\\
{\bf a}({\bf x}_R,\omega) &\approx& {\bf v}^-({\bf x}_R,\omega).\label{eq316}
\end{eqnarray}
The approximation signs signify that evanescent waves are neglected at ${{\partial\mathbb{D}}_R}$. 
Substitution of equations (\ref{eq315}) and (\ref{eq316}) into equation (\ref{eq318f}) gives equation (\ref{eq318}).

\subsection{Derivation of the modified elastodynamic Green's function}\label{AppB2}

We derive a modified elastodynamic Green's function ${\bf \Gamma}({\bf x},{\bf x}_S,\omega)$ (with $x_{3,S}=x_{3,R}-\epsilon$), such that
for ${\bf x}$ at ${{\partial\mathbb{D}}_R}$, i.e., just below the source level, the downgoing part of ${\bf \Gamma}({\bf x},{\bf x}_S,\omega)$ obeys equation (\ref{eq4Gag}), i.e.,
\begin{eqnarray}
\lim_{x_{3}\downarrow x_{3,S}}{\bf \Gamma}^+ ({\bf x},{\bf x}_S,\omega)={\bf I}\delta({\bf x}_{\rm H}-{\bf x}_{{\rm H},S}).\label{eq4Gagag}
\end{eqnarray}
To this end we first investigate the properties of the downgoing part of ${\bf G}({\bf x},{\bf x}_S,\omega)$ defined in equations (\ref{eqweedynG}) and (\ref{eq319}), just below the source level.
Consider  the inverse of equation (\ref{eq305})
\begin{eqnarray}
&&\hspace{-0.7cm}
\begin{pmatrix} \tilde{\bf v}^+\\\tilde{\bf v}^- \end{pmatrix}=
\begin{pmatrix}  -\tilde{\bf \Delta}^{-1}\tilde{\bf D}^-& \tilde{\bf \Delta}^{-1}\\
\tilde{\bf \Delta}^{-1}\tilde{\bf D}^+ & -\tilde{\bf \Delta}^{-1} \end{pmatrix}
\begin{pmatrix} \tilde{\bf v} \\-\tilde{\mbox{\boldmath $\tau$}}_3\end{pmatrix},
\quad\mbox{for}\quad x_3\le x_{3,R}\label{eqB16}
\end{eqnarray}
with
\begin{eqnarray}
\tilde{\bf \Delta}=\tilde{\bf D}^+- \tilde{\bf D}^-.
\end{eqnarray}
The upper-right matrix in equation (\ref{eqB16}), $\tilde{\bf \Delta}^{-1}$, gives the relation between $-\tilde{\mbox{\boldmath $\tau$}}_3$ and the downgoing velocity vector $\tilde{\bf v}^+$.
The same matrix transforms a unit force source in a homogeneous half-space into the downgoing part of the Green's function just below this source, hence
\begin{eqnarray}
&&\hspace{-.5cm}\lim_{x_3\downarrow x_{3,S}}\tilde {\bf G}^+({\bf s},x_3,{\bf 0},x_{3,S},\omega)=
\tilde{\bf \Delta}^{-1}({\bf s})=\nonumber\\
&&\hspace{-.5cm}\frac{1}{2\rho_0}\begin{pmatrix}
\frac{s_1^2}{s_3^P }+\Bigl(\frac{1}{c_S^2}-s_1^2\Bigr)\frac{1}{s_3^S} & \Bigl(\frac{1}{s_3^P }-\frac{1}{s_3^S}\Bigr)s_1s_2 & 0\\
\Bigl(\frac{1}{s_3^P }-\frac{1}{s_3^S}\Bigr)s_1s_2 & \frac{s_2^2}{s_3^P }+\Bigl(\frac{1}{c_S^2}-s_2^2\Bigr)\frac{1}{s_3^S} & 0\\
0 & 0 & s_3^P +\frac{{\bf s}\cdot{\bf s}}{s_3^S}
\end{pmatrix}.\nonumber\\
&&\label{eq325}
\end{eqnarray}
In equation (\ref{eq325}) the source is located at $({\bf 0}, x_{3,S})$.
Next, we consider ${\bf G}({\bf x},{\bf x}_S,\omega)$ for a laterally shifted source position $({\bf x}_{{\rm H},S}, x_{3,S})$.
Applying a spatial Fourier transform along the horizontal source coordinate ${\bf x}_{{\rm H},S}$, using equation (\ref{eq329}) with ${\bf x}_{{\rm H},R}$ replaced by ${\bf x}_{{\rm H},S}$, 
yields  $\tilde{\bf G}({\bf x},{\bf s},x_{3,S},\omega)$. For the downgoing part just below the source we obtain a phase-shifted version of 
the Green's function of equation (\ref{eq325}), according to
\begin{eqnarray}
\lim_{x_3\downarrow x_{3,S}}\tilde{\bf G}^+({\bf x},{\bf s},x_{3,S},\omega)
&=&\tilde{\bf \Delta}^{-1}({\bf s})\exp\{i\omega{\bf s}\cdot{\bf x}_{\rm H}\}.\label{eq330k}
\end{eqnarray}
Comparing this with 
the desired condition of equation (\ref{eq4Gagag}) suggests to define the modified Green's function
(for arbitrary ${\bf x}$) as
\begin{eqnarray}
&&\tilde{\bf \Gamma}({\bf x},{\bf s},x_{3,S},\omega)=\tilde{\bf G}({\bf x},{\bf s},x_{3,S},\omega)\tilde{\bf \Delta}({\bf s}),\label{eq331}
\end{eqnarray}
such that
\begin{eqnarray}
\lim_{x_{3}\downarrow x_{3,S}}\tilde{\bf \Gamma}^+ ({\bf x},{\bf s},x_{3,S},\omega)
={\bf I}\exp\{i\omega{\bf s}\cdot{\bf x}_{\rm H}\}.\label{eq544}
\end{eqnarray}
The inverse Fourier transform from ${\bf s}$ to ${\bf x}_{{\rm H},S}$ gives indeed equation (\ref{eq4Gagag}). 

We define the reflection response $\tilde{\bf R}({\bf x}_R,{\bf s},x_{3,S},\omega)$ of the medium below ${{\partial\mathbb{D}}_R}$ 
as the upgoing part of the modified Green's function $\tilde{\bf \Gamma}({\bf x}_R,{\bf s},x_{3,S},\omega)$,  with ${\bf x}_R$ at ${{\partial\mathbb{D}}_R}$, hence
\begin{eqnarray}
\tilde{\bf R}({\bf x}_R,{\bf s},x_{3,S},\omega)&=&\tilde{\bf \Gamma}^-({\bf x}_R,{\bf s},x_{3,S},\omega),\label{eq3456}
\end{eqnarray}
or, using equation (\ref{eq331}),
\begin{eqnarray}
\tilde{\bf R}({\bf x}_R,{\bf s},x_{3,S},\omega)&=&\tilde{\bf G}^-({\bf x}_R,{\bf s},x_{3,S},\omega)\tilde{\bf \Delta}({\bf s})\nonumber\\
&=&\tilde{\bf G}^{\rm s}({\bf x}_R,{\bf s},x_{3,S},\omega)\tilde{\bf \Delta}({\bf s}),\label{eq5432}
\end{eqnarray}
where superscript ${\rm s}$ stands for scattered.
The inverse Fourier transform of equation (\ref{eq3456}) from ${\bf s}$ to ${\bf x}_{{\rm H},S}$ yields equation (\ref{eq4Rag}).

\subsection{Derivation of the  representation of equation (\ref{eq349})}\label{AppB3}

 To obtain a representation for ${\bf G}({\bf x},{\bf x}_S,\omega)$ we start by  transforming all terms in equation (\ref{eq339}) along ${\bf x}_{{\rm H},S}$, 
 using equation (\ref{eq329}), with ${\bf x}_{{\rm H},R}$ replaced by ${\bf x}_{{\rm H},S}$, hence
\begin{eqnarray}
\tilde{\bf \Gamma}({\bf x},{\bf s},x_{3,S},\omega)&=&\int_{{{\partial\mathbb{D}}_R}} {\bf F}({\bf x},{\bf x}_R,\omega)\tilde{\bf R}({\bf x}_R,{\bf s},x_{3,S},\omega){\rm d}{\bf x}_R 
+\tilde{\bf F}^*({\bf x},-{\bf s},x_{3,S},\omega),\nonumber\\
&&\hspace{7cm}\mbox{for}\quad x_3 \ge x_{3,R}.
\label{eq340}
\end{eqnarray}
We introduce a modified focusing function $\tilde{\bf f}({\bf x},{\bf s},x_{3,S},\omega)$ via
\begin{eqnarray}
\tilde{\bf F}({\bf x},{\bf s},x_{3,S},\omega)=\tilde{\bf f}({\bf x},{\bf s},x_{3,S},\omega)\tilde{\bf \Delta}({\bf s}).\label{eq341z}
\end{eqnarray}
According to equation (\ref{eq325}) we have  for propagating waves (i.e., for  ${\bf s}\cdot{\bf s}\le 1/c_P^2$)
\begin{eqnarray}
\tilde{\bf \Delta}({\bf s})=\tilde{\bf \Delta}(-{\bf s})=\tilde{\bf \Delta}^*({\bf s})=\tilde{\bf \Delta}^t({\bf s}).\label{eqdsym}
\end{eqnarray}
Hence, for  $\tilde{\bf F}^*({\bf x},-{\bf s},x_{3,S},\omega)$ we obtain
\begin{eqnarray}
&&\tilde{\bf F}^*({\bf x},-{\bf s},x_{3,S},\omega)=\tilde{\bf f}^*({\bf x},-{\bf s},x_{3,S},\omega)\tilde{\bf \Delta}({\bf s}).\label{eq341}
\end{eqnarray}
Multiplying all terms in equation (\ref{eq340})  from the right by $\tilde{\bf \Delta}^{-1}({\bf s})$, 
using equations (\ref{eq331}),  (\ref{eq5432}) and (\ref{eq341}), and transforming the resulting expression back from ${\bf s}$ to ${\bf x}_{{\rm H},S}$ gives
\begin{eqnarray}
{\bf G}({\bf x},{\bf x}_S,\omega)&=&\int_{{{\partial\mathbb{D}}_R}} {\bf F}({\bf x},{\bf x}_R,\omega){\bf G}^{\rm s}({\bf x}_R,{\bf x}_S,\omega){\rm d}{\bf x}_R 
+{\bf f}^*({\bf x},{\bf x}_S,\omega), \quad\mbox{for}\quad x_3 \ge x_{3,R}.\label{eq342}
\end{eqnarray}
We  modify the integral step by step. First we use source-receiver reciprocity for the scattered Green's function ${\bf G}^{\rm s}({\bf x}_R,{\bf x}_S,\omega)$ and
we  apply Parseval's theorem. We thus obtain for the integral in equation (\ref{eq342})
\begin{eqnarray}
\frac{\omega^2}{4\pi^2}\int_{\mathbb{R}^2} \tilde{\bf F}({\bf x},-{\bf s},x_{3,R},\omega)\{\tilde{\bf G}^{\rm s}({\bf x}_S,{\bf s},x_{3,R},\omega)\}^t{\rm d}{\bf s}.\label{eq344}
\end{eqnarray}
Substituting equation (\ref{eq341z}), using equation (\ref{eqdsym}), gives
\begin{eqnarray}
\frac{\omega^2}{4\pi^2}\int_{\mathbb{R}^2} \tilde{\bf f}({\bf x},-{\bf s},x_{3,R},\omega)\{\tilde{\bf G}^{\rm s}({\bf x}_S,{\bf s},x_{3,R},\omega)\tilde{\bf \Delta}({\bf s})\}^t{\rm d}{\bf s}.\label{eq346}
\end{eqnarray}
Using equation (\ref{eq5432}) this gives
\begin{eqnarray}
\frac{\omega^2}{4\pi^2}\int_{\mathbb{R}^2} \tilde{\bf f}({\bf x},-{\bf s},x_{3,R},\omega)\{\tilde{\bf R}({\bf x}_S,{\bf s},x_{3,R},\omega)\}^t{\rm d}{\bf s}.\label{eq347}
\end{eqnarray}
Applying  Parseval's theorem again and inserting the resulting integral in equation (\ref{eq342}) yields equation (\ref{eq349}).
It  has been derived without applying decomposition in the lower half-space, but it excludes the contribution from waves that are evanescent at ${{\partial\mathbb{D}}_R}$.

\end{spacing}
\end{document}